\documentclass[longauth]{aa} 
\usepackage{graphicx}
\usepackage{natbib}
\usepackage{balance}
\bibpunct{(}{)}{;}{a}{}{,} 

\usepackage{txfonts}
\def\gapprox{\;\rlap{\lower 2.5pt            
 \hbox{$\sim$}}\raise 1.5pt\hbox{$>$}\;}
\def\lapprox{\;\rlap{\lower 2.5pt            
 \hbox{$\sim$}}\raise 1.5pt\hbox{$<$}\;}

\begin{document}

\title{SPHERE/ZIMPOL observations of the symbiotic
  system R Aquarii\thanks{Based on observations collected at the
    European Southern Observatory under ESO programmes
    60.A-9249, 097.D-0516, 0102.D-0501, and 1104.C-0416.}}
   \subtitle{II. Dynamics of the dust in the Mira wind and the H$\alpha$
      gas in the jet}

   \author{
       H.M.~Schmid\inst{\ref{instch1}} 
     \and E.~Lagadec\inst{\ref{instf5}}
     \and J.~Milli\inst{\ref{instf1}}        
     \and C.~Tschudi\inst{\ref{instch1}} 
     \and J.~Ma\inst{\ref{instf1}}
     \and N.~Engler\inst{\ref{instch1}}     
   \and J.L.~Beuzit\inst{\ref{instf3}}
   \and A.~Boccaletti\inst{\ref{instf4}}      
   \and G.~Chauvin\inst{\ref{instd1}}      
   \and R.~Claudi\inst{\ref{insti1}} 
   \and S.~Desidera\inst{\ref{insti1}}
   \and K.~Dohlen\inst{\ref{instf3}}        
   \and M.~Feldt\inst{\ref{instd1}}      
   \and R.~Gratton\inst{\ref{insti1}}
   \and Th.~Henning\inst{\ref{instd1}}      
   \and M.~Kasper\inst{\ref{insteso1}}   
   \and M.~Langlois\inst{\ref{instf7}}      
   \and F.~M\'enard\inst{\ref{instf1}}         
    \and D.~Mouillet\inst{\ref{instf1}}
   \and A.~Vigan\inst{\ref{instf3}}        
          }

\institute{
ETH Zurich, Institute for Particle Physics and Astrophysics,
Wolfgang-Pauli-Strasse 27, 
CH-8093 Zurich, Switzerland\label{instch1}
\and
Laboratoire Lagrange, UMR7293, Universit\'{e} de Nice Sophia-Antipolis, 
CNRS, Observatoire de la C\^{o}te d'Azur, Boulevard de l'Observatoire, 
06304 Nice, Cedex 4, France\label{instf5}
\and
Universit\'{e} Grenoble Alpes, CNRS, IPAG, 38000 Grenoble, France\label{instf1}
\and
Aix Marseille Univ, CNRS, LAM, Laboratoire d'Astrophysique de Marseille,
Marseille, France\label{instf3}
\and
LESIA, Observatoire de Paris, PSL Research University, 
CNRS, Sorbonne Universités, UPMC Univ. Paris 06, Univ. Paris Diderot, 
Sorbonne Paris Cité, 5 place Jules Janssen, 92195 Meudon, France\label{instf4}
\and
Max-Planck-Institut f\"{u}r Astronomie, K\"{o}nigstuhl 17, 69117
Heidelberg, Germany\label{instd1}
\and
INAF – Osservatorio Astronomico di Padova, Vicolo
dell’Osservatorio 5, 35122 Padova, Italy\label{insti1}
\and
European Southern Observatory, Karl Schwarzschild St, 2, 85748
Garching, Germany\label{insteso1}
\and
Centre de Recherche Astrophysique de Lyon, CNRS/ENSL
Universit\'{e} Lyon 1, 9 av. Ch. Andr\'{e}, 69561 Saint-Genis-Laval,
France\label{instf7}
}

\abstract
    {R~Aqr is a resolved symbiotic Mira binary and a
      prototype target for the study of complex hydrodynamic processes
      including dust-driven mass loss, mass transfer, accretion,
      and jet outflows and the formation of strongly asymmetric
      circumstellar structures.}
    {With the aim of gaining a better understanding of
      the complex hydrodynamics of R~Aqr, we conducted an
      investigation based on new high-resolution imaging of the H$\alpha$
      emission from the ionized gas and of the polarization from
      scattering dust.}
    {We analyzed SPHERE/ZIMPOL H$\alpha$ and polarization maps 
      with a resolution of $\approx 25$ mas (5~AU)
      for the innermost region -- $r\lapprox 2''$ ($\lapprox 400$~AU) -- of
      the R~Aqr system.
      Data for four epochs were taken covering 6.7~years, including
      the dust eclipse phase around 2021,
     providing the angular motion and structural changes for the
     H$\alpha$ and dust clouds, which are measured and described in detail.
     The results are compared with literature data and put into context
     with respect to models for AGB star binaries.}
    {The central H$\alpha$ emission traces the
      orbital motion of the hot component.
      The circumstellar polarization from scattering
      dust shows interaction effects near the binary ($\lapprox 60$~AU),
      including a flattened distribution aligned with the orbital plane,
      corotating dust, and a reduction of the scattering signal during
      the dust eclipse phase. The polarization shows for $\gapprox 60$~AU
      a steady outflow with about a dozen dust clouds,
      a few of which have a comet-like core and tail features.
      The dust velocity is about $20~{\rm km\,s}^{-1}$,
      and this is faster than for the CO gas measured with ALMA.
      Several prominent H$\alpha$ clouds near the binary ($\lapprox 200$~AU)
      also show scattering polarization because they are just
      partly ionized dust clouds from the Mira wind. We
      also found small faster-moving transient H$\alpha$ clouds, pointing
      to outflow instabilities such as those in clumpy winds of hot stars. The
      observed temporal changes of the H$\alpha$ emission, in particular the
      overall angular shift during the dust eclipse event, can
      be explained by strong illumination changes from the hot component.       
      At a larger separation ($\gapprox 500~$AU), the H$\alpha$ clouds move
      faster ($>100~{\rm km\,s}^{-1}$), in agreement with previous studies.
      This acceleration in the outflow is probably a result of gas
      heating by photoionization and shocks in the clumpy outflow, and
      the produced overpressure gives rise to the two well-known
      jet-like features expanding in polar directions.}
      {Monitoring of the inner regions of R~Aqr shows
      the motion and evolution of different structures
      in the dusty Mira wind and the ionized outflow of the hot component,
      and these features provide unambiguous information about the
      hydrodynamics of the
      circumstellar material. Such observations are very
      useful to clarify the complex phenomena taking place in
      interacting binaries with asymptotic giant branch stars.}
    

\authorrunning{H.M. Schmid et al.}

\titlerunning{SPHERE/ZIMPOL observations of the symbiotic system R~Aqr}

\maketitle

\section{Introduction} \label{Introduction}
The symbiotic Mira star R~Aqr shows very different types of
binary interactions, including asymmetric mass loss, mass transfer,
accretion, and jet features from the compact companion, and a very extended
ionized nebula. R~Aqr has an orbital period of about 43~years
\citep{Gromadzki09,Alcolea23}, and the Mira wind has a pulsation period
of 387 days. R Aqr is one of the nearest symbiotic
stars, and its circumstellar outflow has been studied in
great detail with various types of high-resolution imaging techniques. 
The central binary was resolved in 2014 with a separation of 
45~mas \citep[][hereafter Paper I]{Schmid17}, and the diameter
of the Mira is about 15~mas \citep[e.g.,][]{Zhao-Geisler12}.
The distance to R~Aqr is rather uncertain, and estimates range from 
170~pc to more than 300~pc \citep[see, e.g.,][]{Liimets25}. 
In this work, we adopt $d = 218$~pc from Very Long Baseline
Interferometry (VLBI) parallax
measurements of SiO masers \citep{Min14}, and this yields
3.3~AU for the diameter of the Mira and more than 12~AU for the binary
separation. 

The extended nebulosity, with a diameter of about 120~arcsec in the
east-west direction and 60~arcsec in the north-south direction, was already studied 
by \cite{Lampland22} \citep[see also][]{Hollis99a}, and the
detection of a stellar jet \citep{Herbig80} triggered
many detailed studies on the nebular structure and
the dynamics with observations providing a continuously increasing
data quality with time
\citep[e.g.,][]{Sopka82,Solf85,Hollis99b,Liimets18,Santamaria24}. 
R~Aqr was mapped with 
a high spatial resolution of $\lapprox 100~$mas in emission line filters
using the Hubble Space Telescope (HST)
\citep[e.g.,][]{Paresce94,Hollis97a,Melnikov18,Huang23}
or ground-based adaptive optics observations
\citep{Schmid17,Lucas24} and speckle interferometry \citep{Liimets25},
radio interferometry of the free-free
emission from ionized jet clouds \citep{Dougherty95,Cotton21}
and of maser lines from the inner Mira wind
\citep{Bobholz97,Hollis01,Ragland08,Kamohara10},
near-infrared interferometry of the Mira photosphere \citep{Ragland08},
including the surrounding dust emission \citep{Tuthill00,Zhao-Geisler12},
and ALMA high-resolution observations of the molecular outflow of CO
\citep{Ramstedt18,Bujarrabal18,Bujarrabal21}, and maps of other lines
and the millimeter-continuum \citep{GomezG24}.
In addition, there exist for R~Aqr extensive spectroscopic
data for the visual \citep[e.g.,][]{Wallerstein80,Hollis91}, the UV
\citep[e.g.,][]{Kafatos86,Meier95}, the X-ray
\citep[e.g.,][]{Kellogg07,Toala22,Sacchi24}, and the
infrared ranges \citep[e.g.,][]{Hinkle22,Sankrit22}. 

It is widely accepted that the hot component is an accreting
white dwarf responsible for the collimated outflow
and the ionized nebula around R~Aqr. Different models have been put
forward for the ionization of the
gas, such as photoionization by the radiation from a hot white dwarf
with $T_{\rm eff}\approx 55\,000$~K \citep[e.g.,][]{Kaler81}, the
emission from the accretion shock near the white dwarf, or collisional
ionization by shocks from the supersonic gas outflow, or a combination
of these processes \citep[e.g.,][]{Burgarella92,Huang23}

The orbital period of about 44 years for the R~Aqr binary was
first inferred from periodic phases of reduced brightness,
$\Delta {V}_{\rm vis}\approx 2^m$,
for the Mira observed around 1890, 1933, and 1977 \citep{Willson81}
because of the obscuration by a large dust cloud moving with
the hot component passing periodically in front of the Mira
\citep{Whitelock83}. Another dust obscuration event
occurred more recently between 2019 and 2023 \citep[e.g.,][]{Omelian25}, and this
interesting phase is covered by the VLT/SPHERE observations
presented in this work. Fortunately, there exist many other
types of observations for this event, including extensive photometric
and spectroscopic monitoring in the visual from AAVSO,
X-ray observations \citep{Toala22,Sacchi24}, HST emission line observations,
IR spectroscopy \citep{Hinkle22}, mid-IR and far-IR dust emission measurements
\citep{Omelian25,Sankrit22}, and
millimeter and centimeter wave observations of the CO gas \citep{Bujarrabal21} and
the continuum \citep{Cotton21,GomezG24}, and there is probably
more data yet to be published. The occultation coincides roughly
with the periastron phase of the
binary orbit with an eccentricity of $\epsilon \approx 0.2-0.5$,
and the estimated physical separation of the two stars for this phase is 
about 10~AU only \citep[e.g.,][]{Gromadzki09,Alcolea23}.

This work presents for the central region of R~Aqr
high-resolution differential imaging for the H$\alpha$ emission
and polarimetric imaging of the scattered radiation from 
the circumstellar dust for four epochs between
2014 and 2021 taken with the visual camera ZIMPOL of the 
SPHERE ``VLT planet finder'' instrument. The H$\alpha$ maps from
2014 were already described in Paper~I, but the three additional
epochs now allow for study of the evolution of the H$\alpha$ emission
with time. These data are complemented with archival H$\alpha$
observations of R~Aqr from HST covering five epochs between 2014 and 2023.
The imaging polarimetry provides a detailed mapping of the changes of the 
circumstellar dust simultaneously with the H$\alpha$ observations
and also covers the dust obscuration event for the Mira in R Aqr.
The presented imaging polarimetry are the first data of their kind
for this system.

The following section and Appendix\ref{ZIMPOLdata}
provide technical details of the observations and the system parameters of
R~Aqr for the four observing epochs.
In Section~\ref{HotHalpha} we
describe the motion and changes of the H$\alpha$ emission associated
with the hot companion.
The focus of Sect.~\ref{SectInteract} is on the H$\alpha$ and dust
structure in the interaction region near the binary $r\lapprox 0.3''$.
The regions further out, at separations $r\approx 0.3''$ to $2''$,
are dominated by a radial outflow, and the observed changes are
described in Sect.~\ref{SectOutflow}. 
In Section~\ref{SectPMall} we summarize the outflow properties
for R~Aqr, and in Sect.~\ref{Discussion}
the observations are put into context with other
observations for R~Aqr and with binary models for asymptotic giant
branch (AGB) stars with the aim of better understanding
the hydrodynamics of the circumstellar gas.

\section{Observations and data reduction}
\label{SectObs}

\subsection{SPHERE/ZIMPOL observations}

R~Aqr data were taken with the ESO VLT telescope using the
SPHERE/ZIMPOL instrument, a high-resolution imager operating in the visual
using extreme adaptive optics, and differential imaging techniques for
high contrast observations \citep{Beuzit18,Schmid18}. ZIMPOL provides a
field of view of about $3.6'' \times 3.6''$ with a pixel scale of 3.6~mas/pix
around the central star which is used as AO guide star.
The SPHERE AO system provides in the I-band an angular resolution of
about 25 mas, with a Strehl ratio up to 50~\% under optimal
observing conditions.
R~Aqr was used in 2014 as test target for the commissioning for
intensity imaging and polarimetric observations with the ZIMPOL imager.
A detailed description of maps
for H$\alpha$ and other emission lines from these observations
are given in Paper~I,
while the performance of Stokes $Q$ and $U$ polarization measurements
for the scattered light is described in \citet{Schmid18}.

Subsequently, we reobserved R~Aqr as part of the SPHERE
guaranteed time observing program in 2016, 2018 and 2021
in visitor mode in order to follow temporal changes
including the predicted dust obscuration event.
In each run, spectral differential imaging with simultaneous
observations in the narrow H$\alpha$ filter (N\_Ha) and the
nearby continuum filter (Cnt\_Ha) for the mapping of the ionized gas
were obtained. In addition, we obtained 
polarimetric imaging in the I-band with short,
non-saturated observations for the investigation of the immediate
surrounding $r\lapprox 0.2''$ of the Mira. Deeper, coronagraphic,
observations were taken with longer exposures to measure the
fainter clouds at larger separation $r> 0.2''$.
Three different I-band filters were used, I\_PRIM, N\_I and Cnt820
with practically the same central wavelength of 790~nm, 817~nm and
817~nm, respectively. The filter width has little
influence on the results 
because dust scattering properties are very similar for small
wavelength differences. More technical details of the
presented ZIMPOL data are given in Appendix~\ref{ZIMPOLdata}.

Polarimetry in the V and R bands was also taken for the investigation
of the dust properties from multiwavelength measurements of the brightest
dust scattering region. However, these data are much less sensitive, 
and they are not used for this study focussing on the
overall mapping of the circumstellar region.

\begin{figure}
\includegraphics[trim=0.cm 0.cm 1.5cm 1.5cm,clip,width=8.5cm]{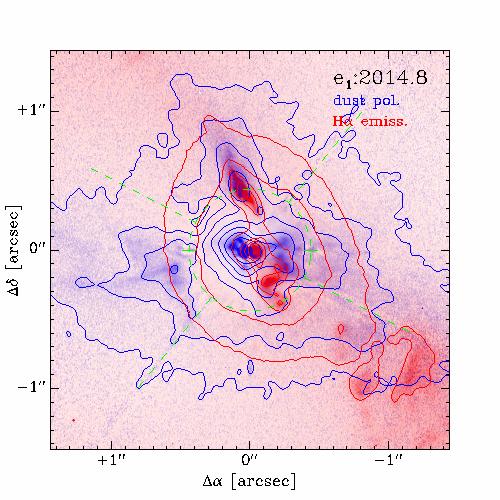}  
\caption{H$\alpha$ intensity $I_{{\rm H}\alpha}$ (red) and I-band 
  polarized intensity $Q_{\phi}^{\rm ex}$ (blue) for the R~Aqr observations
  from 2014. The contours have been smoothed and are given
  for increments of a factor of two for the $I_{{\rm H}\alpha}$ and a factor of
  3.16 for $Q_{\phi}^{\rm ex}$. The colors have been scaled with a separation of
  $\propto r^\alpha$ with $\alpha\approx 2$.
  Dashed green lines mark regions affected by PSF features.}
\label{HaDust14}
\end{figure}

\paragraph{An example for the resulting maps}
Differential maps for the H$\alpha$ emission
$I_{{\rm H}\alpha}(\alpha,\delta)$ and the extended (or resolved)
azimuthal polarization from the scattering dust $Q_\phi^{\rm ex}(\alpha,\delta)$
are shown as example in Fig.~\ref{HaDust14}
for the observations from epoch e$_1$ taken for 2014.8.
H$\alpha$ is plotted in red
and $Q_\phi^{\rm ex}$ in blue, and the colors were scaled 
with radial functions of the type $\propto r^\alpha$ with $\alpha\approx 2$
to highlight the structures of the very bright clouds near
the central binary and the much fainter structures at
larger separation as described in Appendix~\ref{Maps234}.
The seven contours for the dust give the flux levels for
increments of a factor of 3.16 covering a range of 1000,
while the increments for H$\alpha$ are a factor of two and the covered
range is 32 for six contours. 

The map shows clearly the red H$\alpha$ clouds in the two narrow wedges
with a position angle of about $15^\circ$ (measured N over E) and
$225^\circ$ representing the collimated outflows of R~Aqr. The
blue dust scattering clouds have predominantly an E-W orientation aligned
roughly with the expected orientation of the orbital plane. There
is a bright dust cloud about $0.4''$ north of the binary, which
coincides with a strong H$\alpha$ emission structure. Because 
H$\alpha$ and $Q_\phi$ are observed with the same instrument, the 
spatial resolution and the relative astrometry of features for the
ionized gas and the polarized dust emission match perfectly with
an accuracy of about $\pm 2$~mas. Similar maps for the epochs
e$_2$, e$_3$, and e$_4$ are presented in the Appendix (Sect.~\ref{Maps234}).
Obviously, the circumstellar material shows complex structures
evolving significantly with time.

\subsection{Data reduction}
\label{SectReduction}
The data reduction was performed with the SZ-software
developed at the ETH Zurich which includes standard steps, such as 
bias frame subtraction, outlier pixel removal with a median filter,
and flat fielding. In addition we applied for polarimetric data
a polarimetric efficiency calibration and
a correction for the polarimetric beam shift as described
in \cite{Schmid18}.

H$\alpha$ differential images are obtained with simultaneous
observations in the N\_Ha filter with camera1 of ZIMPOL
and a CntHa filter with camera2. The two frames show
a very similar structure for the point spread function (PSF)
and the speckle pattern, and 
the subtraction of a scaled image $I_{\rm CntHa}$ from
the corresponding image $I_{\rm N\_Ha}$ yields
the differential flux $I_{{\rm H}\alpha}$ without the Mira
continuum according to
\begin{equation}
  I_{{\rm H}\alpha}(\alpha,\delta) = I_{\rm N\_Ha}(\alpha,\delta)
    - a \cdot I_{\rm CntHa}(\alpha,\delta)\,.
\label{EqDiffima}  
\end{equation}
The scaling factor $a$ is derived so that the
flux of the point-like peak of the Mira is canceled in the   
difference image $I_{{\rm H}\alpha}$. The H$\alpha$
emission at larger separation $\gapprox 0.2''$
is hardly affected by this procedure and the subtraction is
straight forward. But, near the very bright Mira
already small differences in the PSFs and the speckle patterns
between $I_{\rm N\_Ha}$ and $I_{\rm CntHa}$ produce
relatively strong residuals in the $I_{{\rm H}\alpha}$ image.
The best results are obtained by selecting only high quality
CntHa/N\_Ha image pairs and aligning them with subpixel precision
to obtain minimal
residuals at the location of the Mira as described for the
2014 data in Paper~I. It should be noted that this procedure
subtracts also an unresolved H$\alpha$ emission component
coinciding with the point-like Mira variable, such as a possible H$\alpha$
emission produced by shocks in the pulsating atmosphere of
the Mira star \citep[e.g.,][]{Fox84,Boyd26}.

The same detector integration time (DIT) of
50~s was used for the CntHa/H\_Ha filter observations for the four
epochs, and therefore the count signals $I_{{\rm H}\alpha}(\alpha,\delta)$
for the different epochs are on a similar flux scale, with an
uncertainty of about a factor of two. A more accurate flux calibration
of the ZIMPOL H$\alpha$ data would require the modeling of PSF
convolution effects as presented in Paper~I for the epoch 2014.8
and additional corrections factors for the observations of 2016.6 and 2021.5
taken under non-photometric conditions. Fortunately, the
H$\alpha$ from HST described in Sect.~\ref{SectHSTdata}
provide an independent check for the H$\alpha$ maps obtained
with SPHERE/ZIMPOL.

The polarimetric data reduction uses a double-difference
method for cycles of four observations taken with 
different half-wave plate positions. This yield maps for the 
Stokes parameters $Q(\alpha,\delta)$ and $U(\alpha,\delta)$ for
the linear polarization in sky coordinates as primary results. Because of the
very strong brightness contrast, we took short non-saturated
data for the Mira and its immediate surroundings and much deeper
saturated or coronagraphic observations for the scattering polarization
at separations $\gapprox 0.3''$, as described in detail for the
2014 observation in \cite{Schmid18}. We then combined 
the short and deep observations to single
$Q(\alpha,\delta)$ and $U(\alpha,\delta)$ frames covering the entire
field of view.

The obtained polarization maps include offsets of about
$p\approx 0.2$~\% introduced by the telescope polarization and
for some observing epochs also offsets $p>0.2$~\% introduced
by the intrinsic polarization of the dominant Mira emission. 
We applied a polarimetric zero-point correction (zp correction), which
sets the polarization of the central source to zero,  
according to the prescription given in \cite{Schmid26}.
This zp correction accounts for the polarization of the
telescope and the central star introducing offsets
$\langle q_0 \rangle$ and $\langle u_0 \rangle$ according to
\begin{eqnarray}
  Q^{\rm ex}(\alpha,\delta) =
  Q(\alpha,\delta)-\langle q_0 \rangle \, I(\alpha,\delta)\,,\label{EqQex}\\
  U^{\rm ex}(\alpha,\delta) =
  U(\alpha,\delta)-\langle u_0 \rangle \, I(\alpha,\delta)\,,\label{EqUex}
\end{eqnarray}
where $\langle q_0 \rangle = Q_0/I_0$ is the fractional
polarization measured for the central peak and equivalent for
$\langle u_0 \rangle$.
This procedure splits the polarization signal into
an unresolved polarization component for the Mira variable $Q_0$ and
$U_0$ and an extended, spatially resolved polarization component
$Q^{\rm ex}(\alpha,\delta)$ and $U^{\rm ex}(\alpha,\delta)$ for the
circumstellar scattering. From the $Q^{\rm ex}$ and $U^{\rm ex}$ maps, we
derived the azimuthal polarization for the spatially
resolved circumstellar dust scattering: 
\begin{equation}
 Q^{\rm ex}_\phi(\alpha,\delta) = -Q^{\rm ex}(\alpha,\delta)\, \cos(2\phi_{\alpha\delta})
               -U^{\rm ex}(\alpha,\delta)\,\sin(2\phi_{\alpha\delta})\,,
\label{QphiEx}
\end{equation}
where $\phi_{\alpha\delta}$ is the position angle with respect to
the Mira. The $Q_\phi^{\rm ex}$ map serves as an approximation for
the spatially resolved polarized flux ($p\times I$) from the
circumstellar scattering region. The $Q_\phi^{\rm ex}$ signal
is preferred because it is not biased by pixel-to-pixel noise \citep{Schmid06,Schmid26}.

\begin{figure}
\includegraphics[trim=2.5cm 13.2cm 8.6cm 3.0cm,clip,width=8.6cm]{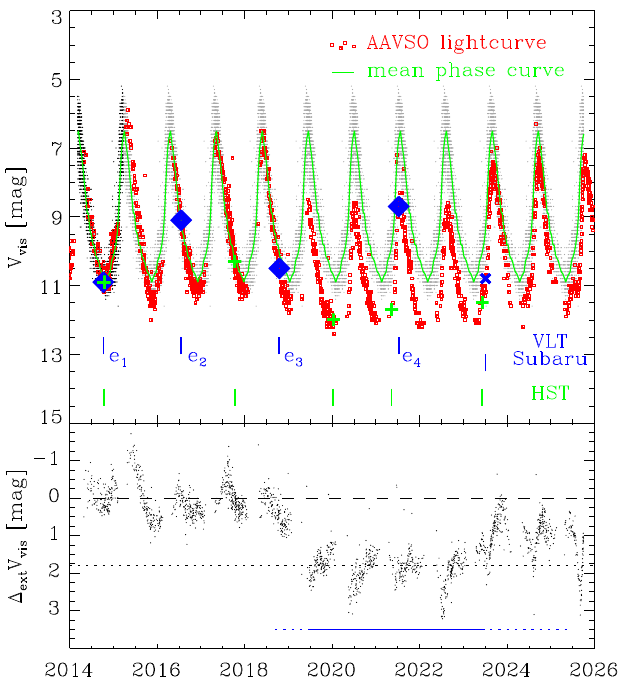}  
\caption{Upper panel: R~Aqr light curve V$_{\rm vis}$ from 2014 to 2025
from the AAVSO database (red squares) plotted together with
the mean pulsation phase light curve ${\rm V}_{\rm max}(\phi)$ for the years
2000 to 2015 (green line). Small gray points are individual measurements
for this phase curve. High-resolution imaging observations are indicated: blue diamonds are for the VLT, the blue cross is for Subaru, and green crosses are
for HST. Lower panel: Difference ${\rm V}_{\rm vis}-{\rm V}_{\rm max}(\phi)$
representing roughly the extra extinction $\Delta_{\rm ext}{\rm V}_{\rm vis}$
during the dust eclipse event lasting from about 2019.5 to 2023.5, as indicated
by the blue line.} 
\label{lightcurve}
\end{figure}

The PSF convolution leads to a smearing
and a cancellation of the polarization signal, and the asymmetric
scattering geometry in R~Aqr produces a systematic
convolution cross talk signal in the $U_\phi^{\rm ex}(\alpha,\delta)$
map. Therefore, the measurable $Q^{\rm ex}_\phi$ signal is
reduced with respect to the intrinsic signal by about a factor of two
for the innermost region $r\lapprox 0.2''$ and by about 30~\% for regions
$\gapprox 0.3''$ \citep[e.g.,][]{Tschudi21}.
In addition, the applied zp correction changes the
angular distribution of the $Q_\phi(\alpha,\delta)$ signal
\citep{Schmid26}. However, we can
assume that the effects are on a similar level for all four ZIMPOL epochs
so that a qualitative description of structural changes
is not significantly affected. Detailed modeling of the instrumental
convolution effects and the zp-correction procedure would be
required to recover accurate quantitative results for the intrinsic
scattering polarization $Q'_\phi(\alpha,\delta)$ near ($r<0.1''$)
the R~Aqr binary \citep{Schmid26}. This is a demanding task and
beyond the scope of this paper.

The central star is highly variable, and the brightness of the scattered radiation changes in step
with these variations. Therefore, we derived relative 
values for the polarization reflectivities, such as $Q_\phi^{\rm ex}/I_\star$,
representing the measured polarization signal
relative to the total intensity of the system $\Sigma I=I_\star$
for an aperture $\Sigma$ with $r=1.5''$ representing the
total brightness of the system.

\subsection{ZIMPOL observing epochs}
The quality of the ZIMPOL data depends significantly
on changing R~Aqr system parameters and observing conditions.
Figure~\ref{lightcurve}
shows the strongly variable, visual light curve V$_{\rm vis}(t)$ of R~Aqr
from the AAVSO database (red points) with a Mira pulsation period
of about $P=384$~days and an amplitude of about $4^m$. Also given
are the VLT, the HST, and the Subaru \citep{Lucas24} observing epochs. 
The H$\alpha$ line and other emissions  
from the ionized nebula are easier to observe when the Mira is faint like
for the VLT epochs 2014.8, and
2018.8, and all the HST observations. On the other side, a bright
Mira produces more dust scattering and the circumstellar polarization
is easier to detect for the 2016.6 and 2021.5 VLT epochs.

The green curve in Fig.~\ref{lightcurve} represents the mean
pulsation phase curve $V(\phi)$ derived from 
the pre-obscuration time 2000 to 2015 according to
${\rm V}_\phi(t)=t_0 + (V(\phi)\,\cdot\,P) \times i$
with $t_0({\rm JD})=2\,451\,350$ and $i=14,..,24$. The used pulsation period
$P=384$~days is derived from the same time interval and it
differs slightly from the period of 387.97~days listed in
the General Catalogue of Variable Stars \citep{Samus17}.
Period changes at this level, and small phase and amplitude changes
are common for light curves of Mira variables
\citep[e.g.,][]{Percy99,Percy19}.

The lower panel shows the difference
$\Delta_{\rm ext} {\rm V}_{\rm vis}={\rm V}_{\rm vis}-{\rm V}_\phi$, and
we use this value as rough estimate for the dust extinction. The
minimum lasts from about 2019.5 to 2023.5 as marked
by the blue line and the Mira brightness is reduced roughly by about
$\Delta_{\rm ext} {\rm V}_{\rm vis}\approx 1.8^m\pm 0.2^m$
or about a factor of five as indicated by the dotted line. There are
quite some deviations from this mean value because the phases
of the individual light curve maxima do not line up exactly with
the maxima of the mean phase curve because of phase variations
mentioned above. Therefore, it is also difficult
to say from the data in Fig.~\ref{lightcurve},
whether the Mira brightness was already affected significantly by dust
obscuration for the VLT epoch 2018.8, for which we derive an extinction
of $\Delta_{\rm ext}{\rm V}_{\rm vis}=0.4^m\pm 0.3^m$. Clearly visible is also,
that the brightness of R~Aqr is still reduced from 2023.5 to 2025.5
by about $\Delta_{\rm ext}{\rm V}_{\rm vis}\approx 1^m$ probably because of
remaining dust extinction during this phase. \cite{Hinkle22} derive
for the I-band a dust extinction of $\Delta_{\rm ext}{\rm I}\approx 2^m$
in 2019 and 2020 from data of the AAVSO archive. The fact,
that the extinctions $\Delta_{\rm ext}{\rm V}_{\rm vis}$ and
$\Delta_{\rm ext}{\rm I}$ are very similar for both bands during the
dust eclipse phase points to an optically thick dust cloud.

The extinction during the last transit around 1976 is investigated
in \citet{Whitelock83}, and they find an
eclipse duration between 4 and 6 years. They measure for the J-band
(1.2~$\mu$m) a strong flux reduction of about
$\Delta_{\rm ext}{\rm J}\approx 2.2^m$ in 1976, and about $0.7^m$ to $1.4^m$
in 1975, 1977, and 1978. For the visual region, they estimated an extinction
of about $\Delta_{\rm ext}V_{\rm vis}\approx 0.7^m$ from the AAVSO light
curve \citep{Mattei79}. This is less than what we derived for the
$\Delta_{\rm ext} {\rm V}_{\rm vis}$ extinction for the 2021
eclipse (Fig.~\ref{lightcurve}), but for a detailed comparison
one should apply the same measuring method for the two events.
Overall, these two eclipses are quite similar. 

The VLT data from 2021.5 were taken near the maximum of the
Mira light curve, but the sensitivity of the
observation for the faint circumstellar H$\alpha$ emission is still
good because the dust extinction of about $1.8^m$ acts
like a natural stellar coronagraph. The situation during the
eclipse is ideal for polarimetry of the unobscured, more distant
scattering regions, because they are strongly illuminated,
while the glare of the Mira is significantly reduced by
the dust obscuration. For these reasons the data from 2021.5
show well the H$\alpha$ emission and the dust scttering despite
the reduced performance of the AO system for
that epoch (Appendix~\ref{Maps234}).
  
\begin{table}
  \caption{R~Aqr observing epochs, photometric parameters, binary separations,
    and PSF characteristics for the VLT data.}  
\label{TabObs}
\begin{tabular}{lcccc}
\noalign{\hrule}
\noalign{\smallskip\hrule\smallskip}
epochs        & e$_1$  & e$_2$ & e$_3$ & e$_4$ \\
\noalign{\smallskip\hrule\smallskip}
year $(t)$    &  2014.8  & 2016.6 & 2018.8 & 2021.5 \\
day of year   & Oct.~11 & July~21 & Oct.~19 & July~11 \\  
MJD $2\,4..\,...$ &  $56\,941$ &  $57\,590$ &  $58\,410$ &  $59\,406$ \\  
\noalign{\smallskip\noindent R~Aqr photometry\tablefootmark{a} \smallskip}
$V_{\rm vis}$ [mag]  & $10.9$ & $9.1 $ & $10.5$ & $8.7$ \\ 
phase $\phi_{\rm max}$\tablefootmark{b}
                 & 0.56  & 0.25   & 0.39  & 0.98  \\  
$\Delta_{\rm ext}{\rm V}_{\rm vis}$ [mag]
                 & $0.0\pm 0.3$ & $0.0\pm 0.3$ & $0.4\pm 0.3$ & $1.8\pm 0.3$ \\
\noalign{\smallskip\noindent offset of hot component from Mira\tablefootmark{c} \smallskip}
$\Delta \alpha_{\rm hot}$ [mas] & $-46$ & $-35$ & $-17$ &  $+18$ \\
$\Delta \delta_{\rm hot}$ [mas] & $+4$  & $-2$  & $-9$  &  $-10$ \\

\noalign{\smallskip \noindent observational conditions \smallskip}
seeing [$''$]   &  0.9-1.3 &  0.35  & 0.5-0.8  &  0.55  \\
transm.         & clear    & clouds &  clear   &  clouds  \\
\noalign{\smallskip \noindent approximative Strehl ratio
  $S_0$ \smallskip}
Cnt\_Ha [\%]   & $\approx$0.05 & 0.15   & 0.11     & $\approx$0.03\tablefootmark{d} \\
I-band  [\%]   & 0.25     & 0.23  & 0.24  &  0.11\tablefootmark{d} \\
\noalign{\smallskip \noindent measured FWHM of I-band PSF  \smallskip}
$d_{\rm meas}$ [mas]     & 28  & 31   & 30     & 40\tablefootmark{d}  \\
\noalign{\smallskip\hrule\smallskip}
\end{tabular}
\tablefoot{
  \tablefoottext{a}{The photometric parameters are based on the AAVSO light curve
  shown in Fig.~\ref{lightcurve};}
  \tablefoottext{b}{pulsation phase, maximum at $\phi_{\rm max}=0$ and $1$;}
  \tablefoottext{c}{based on \cite{Alcolea23}, uncertainty about
    $\pm 15$~mas;} \tablefoottext{d}{AO-loop frequency only 300~Hz.}}
\end{table}

Table~\ref{TabObs} lists for the VLT observations the observing dates
$t$, the brightness and the pulsation phase of R~Aqr, dust extinction
estimates $\Delta_{\rm ext}{\rm V}_{\rm vis}={\rm V}_{\rm vis}(t)-{\rm V_\phi}(t)$,
the binary configuration, and observing conditions.
The orbital position of the hot component relative to the Mira is
inferred from the observed location of prominent
features, which can be associated with the hot component. These are
for example strong H$\alpha$ emission regions \citep{Schmid17,Lucas24},
dust clouds with a strong continuum at 0.8~mm 
\citep{Bujarrabal18,Alcolea23}, or strong free-free emission clouds at 7~mm
\citep{Cotton21}. These positions must also be compatible with
the expected motion of the hot companion according to the RV curve of the
Mira derived from maser emissions \citep{Gromadzki09,Alcolea23}.
These data firmly established that the hot component was in
2014 about 45~mas to the west of the Mira and passed around
2019 roughly in front of the Mira as confirmed by the obscuration event.
An analysis of the orbital motion for R~Aqr is presented in \cite{Alcolea23}
and we interpolate between the given points to derive 
the relative positions of the hot component with respect to the
Mira $\Delta \alpha_{\rm hot}$ and $\Delta \delta_{\rm hot}$
for the VLT observing epochs given in Table~\ref{TabObs}. We adopt
these values to derive the motion the observed
structures considering possible uncertainties at a level
of about $\pm 15$~mas. Future observations will
provide improved orbital positions.

The quality of the imaging observations depends significantly
on the observing conditions and the AO performance,
and these were quite different for the four epochs. For epoch e$_1$ the
seeing was only mediocre, and for e$_2$ and $e_4$ the conditions
were non-photometric because of thin clouds. For epoch e$_4$ the
AO system was run for unclear reasons with an AO-loop
frequency of only 300~Hz instead of 1380~Hz as in epoch e$_2$ and e$_3$,
or 1200~Hz in e$_1$ during commissioning, and therefore
the AO correction was suboptimal. We can roughly estimate the AO performance
with an approximate Strehl ratio parameter $S_0$ and the
full width at half maximum (FWHM) of the PSF given
in Table~\ref{TabObs}. The value $S_0$ 
compares simply the peak flux for the Mira star with the total counts
within an aperture with a diameter of $3''$ \citep{Schmid18}.
The obtained values underestimate the quality of the imaging data,
because the Mira in R~Aqr has an intrinsic diameter of about
$d_\star\approx 12$~mas to 20~mas as measured by near-IR interferometry
\citep{Millan-Gabet05,Ragland08,Zhao-Geisler12}, and it is therefore not
a point source but an extended source for ZIMPOL. In addition,
there is in R~Aqr a strong contribution of circumstellar flux
of about 10~\%, and probably much more 
for epoch e$_4$ when the light from the star is strongly
suppressed. Nonetheless, the $S_0$-value is very useful to
characterize the relative quality of the data for the epochs
e$_1$, e$_2$ and e$_3$. For epoch e$_4$ it is difficult to disentangle
the effects of the reduced AO performance and the strongly enhanced
relative contribution of circumstellar scattering.

A clear hint for the reduced AO performance for epoch e$_4$
provide the measured I-band widths
of the Mira star of about  $d_{\rm meas}=30 (\pm 2)$~mas for epochs
e$_1$, e$_2$ and e$_3$. This is slightly larger than the
typical instrumental PSF of about $d_{\rm PSF}=25 (\pm 3)$~mas \citep{Schmid18}
and compatible with a
marginally extended Mira star with $d_\star\approx 17$~mas
according to $d_{\rm meas} \approx (d_{\rm PSF}^2+d_\star^2)^{1/2}$.
A significantly larger width of about $d_{\rm meas}=40$~mas
is obtained for e$_4$ and this can be attributed to the
slower AO loop frequency used for this epoch.

\begin{table}
  \caption{Archival H$\alpha$ data from HST included in this study.}  
\label{TabHST}
\begin{tabular}{lcccccc}
\noalign{\hrule\smallskip}
epoch  & day of year & V$_{\rm vis}$ & dataset & exp.~time  \\  
\noalign{\smallskip\hrule\smallskip}
2014.8 & Oct. 18  & $10.9\pm 0.2$  & ic9k06010 & 70\,s \\  
2017.8 & Oct. 13  & $10.3\pm 0.2$  & idc501060\tablefootmark{a} & 80\,s \\  
2020.0 & Jan. 13  & $12.0\pm 0.2$  & ie8702010 & 80\,s \\  
2021.4 & May 13  & $11.7\pm 0.3$  & ieao01040 & 20\,s \\  
2023.4 & May 6   & $11.5\pm 0.3$  & iew901080 & 20\,s \\  
\noalign{\smallskip\hrule\smallskip}
\end{tabular}
\tablefoot{The file names for the used dataset are $*$\_drz.fits from the
  HST programs GO-13339 (PI Stute) and GO-14847,
  GO-16055, GO-16312, GO-17222 (all PI Karovska). All images were taken with
  the same HST ACS/UVIS instrument setup using the filter F656N.
  \tablefoottext{a}{Image saturated at the position of the central binary.}
}
\end{table}

\subsection{Archival HST data for the central H$\alpha$ clouds}
\label{SectHSTdata}

There exist extensive HST-data for R~Aqr taken with the
Ultraviolet-Visible (UVIS) channel of the HST Wide Field
Camera 3 (WFC3) targeting many nebular emission lines of
the system with short and long integrations for
6 epochs between 2013 and 2023. Particularly valuable for
this work are the ``short'' HST exposures in
the H$\alpha$ F656N filter taken for 2014.8, 2017.8, 2020.0, 2021.4, and
2023.4 providing additional epochs for the evolution of the
H$\alpha$ clouds. Important is the high photometric
accuracy of HST for checks of the H$\alpha$ brightness in our
AO data.

Table~\ref{TabHST} lists the H$\alpha$ data from HST considered in this
study, and we used the pipeline processed frames from the HST-MAST
archive with a spatial resolution of about 80~mas for 2 pixels.
Only ``short'' exposures without or with only slight
saturation are selected.
\cite{Melnikov18} used the HST data from 2013 and 2014 for a
comprehensive emission line analysis of the collimated outflow
up to $r\approx 30''$. The data from 2017 are described in \cite{Huang23},
and they analyzed in detail the emission lines
for the inner jet $r<3''$ including also previous HST data for proper
motion measurements.
In Paper~I the HST data from 2014 are used to derive gas densities,
and recombination timescales for the ionized clouds as function of $r$.

\section{The H$\alpha$ emission near the hot component}
\label{HotHalpha}

The spatial resolution of SPHERE/ZIMPOL is high enough to investigate
the immediate surroundings of the individual stellar components of the
binary.
In particular, we can extract the H$\alpha$ emission $I_{{\rm H}\alpha}$ for
regions adding in the N\_Ha filter a clear extra flux component
when compared to the strong PSF of the Mira in the
CntHa image (Eq.~\ref{EqDiffima}). This method is described
for the 2014.8 epoch in Paper~I and
the resulting strong H$\alpha$ emission component was associated
with the hot companion in R~Aqr.

\begin{figure}
\includegraphics[trim=0.4cm 1.0cm 1.5cm 1.5cm,clip,width=8.8cm]{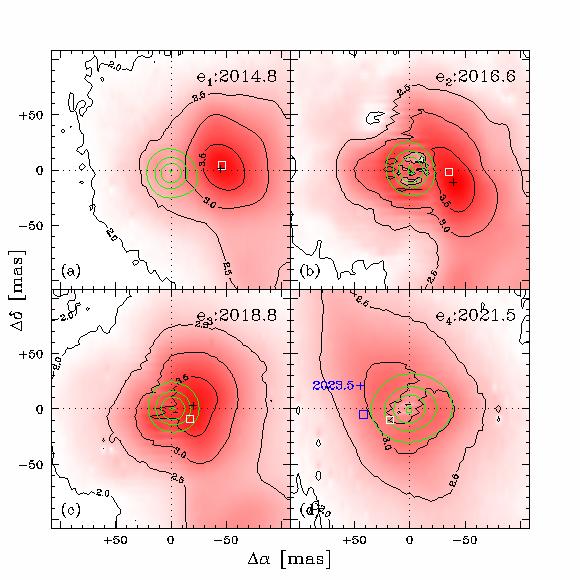}  
\caption{Differential H$\alpha$ maps (black contours and red color) and 
  intensities for relative count levels
  of $c\cdot 10^n$ with $n=2,2.5,3,3.5$. The
  green contours indicate 25\%, 50\%, 75\%, and 99\% of the peak 
  intensity in the CntHa filter. The black crosses
  mark H$\alpha$ cloud peaks, and the white squares are the estimated
  orbital position of the hot component from \cite{Alcolea23}.
  The blue cross in panel (d) gives the H$\alpha$ peak 
  for 2023.5 from \citet{Lucas24}, and the blue square represents the estimated
  position of the hot component for that epoch.}
\label{HaCenter}
\end{figure}

This procedure is applied to the
new SPHERE/ZIMPOL epochs e$_2$, e$_3$, and e$_4$, and
Fig.~\ref{HaCenter} shows the resulting $I_{{\rm H}\alpha}$ maps
together with intensity contours of the Mira obtained
from the simultaneous $I_{\rm CntHa}$ maps (Eq.~\ref{EqDiffima}).
The H$\alpha$ emission shows overall a motion
from west to east and apparently a lower intensity for the
2021.8 dust obscuration phase.

The maps give for the first three
epochs the differential H$\alpha$ count level for a DIT of 50~s and
for the last epoch e$_4$ the registered counts were multiplied by
a factor of 1.33 to account for the reduce AO performance.
Because the Mira was brighter for epochs e$_2$ and
e$_4$ the continuum subtraction introduces in the differential
$I_{{\rm H}\alpha}$ map relatively more noise.
The intrinsic H$\alpha$ flux of
$F_{{\rm H}\alpha}\approx 25\cdot 10^{-12}{\rm erg}{\rm cm}^{-2}{\rm s}^{-1}$
was derived for epoch 2014.8 in Paper~I for the cloud
located 45~mas west of the Mira. This
corresponds to about half of the total H$\alpha$ flux of R~Aqr measured
inside $r\lapprox 1.6''$. This emission can be
explained by an ionized gas nebula with an electron
density of $N_e\approx 4\cdot 10^7\,{\rm cm}^{-3}$
and a mass of $\approx 2\cdot 10^{-8}\,{\rm M}_\odot$ for
a cloud with a radius of about 5~AU which could be ionized
by a hot subdwarf with $T_{\rm eff}\approx 35\,000$~K and
$L=13\,{\rm L}_\odot$ (Paper I). The derived H$\alpha$ flux of
the central cloud for the epochs e$_2$ and e$_3$ is within about
a factor of two at the same level. For epoch e$_4$ the
central H$\alpha$ emission seems to be aligned with the Mira
and the derived $I_{{\rm H}\alpha}$ flux in the region encircled by
the green contours depends strongly on the applied continuum
subtraction. 

The flux peaks for the central $I_{{\rm H}\alpha}$ emission are marked in
Fig.~\ref{HaCenter} with a black cross for the first three epochs.
Clearly there is a motion from west to east and the measured
positions with respect to the Mira are 
$(\Delta\alpha,\Delta\delta) = (-45.1~{\rm mas},+0.4~{\rm mas})$
for epoch e$_1$ adopted from Paper~I, $(-39~{\rm mas},-11~{\rm mas})$
for e$_2$, and $(-20~{\rm mas},+3~{\rm mas})$ for e$_3$, and no
reasonable maximum can be derived for e$_4$.
Also plotted are the expected positions of the hot component
as derived by
\citet{Alcolea23} considering imaging data an RV measurements.

Interesting are the Subaru, SCExAO/Vampires observations 
from \citet{Lucas24} showing for 2023.5 again a strong, point-like
H$\alpha$ source, but located now on the east side of the
Mira at a separation of 50~mas and a position angle of $65^\circ$.
This location is marked in Fig.~\ref{HaCenter}(d) about
25~mas north of the expected hot star position 
according to \citet{Alcolea23}.
This small but significant discrepancy indicates that either the
H$\alpha$ peak does not represent
the position of the hot star well or that the orbital
parameters need to be adjusted. In the SCExAO image the
brightness of the H$\alpha$ source is comparable to
the Mira using a filter such as N\_Ha in ZIMPOL. For this date,
the brightness of the Mira is about $V_{\rm vis}\approx 10.9^m$
practically identical to the e$_1$ epoch. We therefore guess
that the H$\alpha$ flux for 2023.5
is roughly at the level of the e$_1$, e$_2$, and e$_3$ epochs, but this
should be confirmed with a detailed analysis 
of the SCExAO data.

The observed H$\alpha$ structures
are supported by other data. The speckle imaging of
\cite{Liimets25} shows for epoch 2020.9 also an alignment
of H$\alpha$ and Mira continuum emission as for our 2021.5 data.
\cite{Bujarrabal18} find in 2017.9 a strong 0.9~mm continuum feature
in their ALMA map close to the location of our central H$\alpha$ emission
from the 2016.8 epoch. Also interesting are the radio interferometry
data taken with the Jansky Very Large Array (JVLA)
for epoch 2019.6 of the SiO maser emission from near the Mira and
the 7~mm continuum emission from free-free radiation from
\cite{Cotton21}. The central 7~mm emission is 
located only about 10~mas southwest of the Mira, and it is about
five times brighter than the
clouds A$_{\rm SW}$ and C$_{\rm SW}$ also seen in the H$\alpha$ images 
described further below. The 7~mm emission is less affected
by dust extinction and by Mira flux subtraction issues when
compared to H$\alpha$.
The good match between the radio and H$\alpha$ data confirm that
these effects are not strong for the derived H$\alpha$ emission
in Fig.~\ref{HaCenter}(c). The relative location of Mira and
emission nebula is also confirmed by the ALMA continuum and
H30$\alpha$ emission line maps taken at the same time
(2019.7) by \cite{GomezG24}.

The H$\alpha$ emission peaks indicated by crosses show offsets
with respect to the predicted orbital position of
about 10~mas to the south for epoch e$_2$, about 10~mas to the
north for e$_3$, and about 25~mas to the north for the 2023.5 SCExAO
map. Orbital fits producing smaller offsets can certainly
be found but the H$\alpha$ peak positions shows a north-south
wiggling of about $\pm10$~mas not compatible with
the orbital motion. Therefore, the astrometric orbit of the
hot component derived from the H$\alpha$ emission should
account for uncertainties of about $\pm 15$~mas, because the
H$\alpha$ peak flux might be offset from the star position.
Offsets and variations of the H$\alpha$ emission can be 
expected for ionized outflows from accreting stars with inclined accretion
disks, because one of the two polar outflow regions is better visible,
while the other is partly hidden by the opaque disk. Such cases are known
for example for pre-main sequence stars with jets, including DG Tau, FS Tau,
RY Tau, and others \citep[e.g.,]{Eisloeffel98,Garufi19}, and the
geometric morphology could be similar for the H$\alpha$ emission in
R~Aqr. It is also likely that the dust extinction event
can introduce more complexity with respect to the location
of the H$\alpha$ photo center. Observation with higher spatial resolution
or monitoring with high-resolution spectroscopy could reveal
more details about the geometry of the H$\alpha$ emission near the hot star.

\begin{figure}
\includegraphics[trim=0.4cm 1.0cm 1.5cm 1.5cm,clip,width=8.8cm]{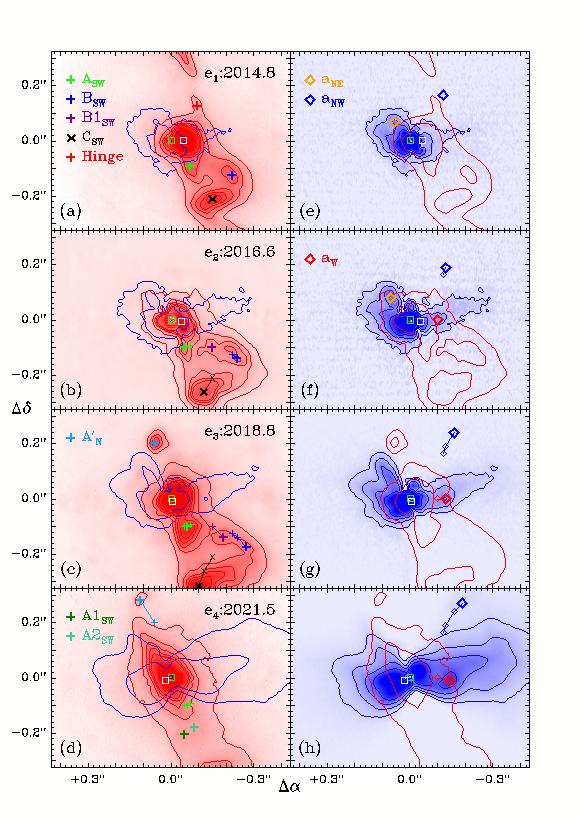}  
\caption{Differential flux $I_{{\rm H}\alpha}$ (left) and I-band
  polarization $Q_\phi^{\rm ex}$ (right) for the center of R~Aqr.
  Relative flux contours for $I_{{\rm H}\alpha}$ are given for levels
  $c\cdot n$ with $n=1,2,3,4,5$ and 
  $c\cdot 2^n$ with $n=0,1,2,3$ for $Q_\phi^{\rm ex}$. 
  Some contours for the polarization are repeated in the H$\alpha$ maps
  and vice versa for comparison.
  H$\alpha$ and $Q_\phi^{\rm ex}$ clouds are identified and replotted with small symbols and lines
  in later epochs to indicate cloud motions. The green and white squares mark the two stellar components.}
\label{HaDustCent}
\end{figure}

\section{Emission from the interaction region}
\label{SectInteract}
The structure of circumstellar material within a few
orbital separations from the center is expected to show
binary interaction effects such as asymmetric mass loss, mass transfer,
accretion, radiation processes, and gravitational acceleration.
Therefore, a detailed investigation of this region may provide information
about these processes, in particular, because we can follow
the temporal evolution of gas structures.

\subsection{H$\alpha$ clouds within 0.3 arcsec or 60~AU}
\label{Hato03}
Overall the H$\alpha$ emission near the central binary from
$r=0.05''$ to $0.3''$ is much stronger on the S-SW side. 
Three main components A$_{\rm SW}$, B$_{\rm SW}$, and C$_{\rm SW}$
where identified in Paper I for epoch e$_1$, and they show quite
strong changes during the covered epoch from
2014 to 2021 (Fig.~\ref{HaDustCent}).

Cloud A$_{\rm SW}$ looks for the first three epochs like an almost
stationary extension to the central H$\alpha$ source at a separation
of about~100 mas to 120~mas from the Mira (Fig.~\ref{HaDustCent}).
For the last epoch e$_4$, there appear two components 
A1$_{\rm SW}$ and A2$_{\rm SW}$ further south at a separation of
about 200~mas. It is difficult to say if these are fast moving
gas clouds in the collimated outflow, which moved from 2018.8 to 2021.5
from near the central source to this position, or whether these are
slow gas clouds which are newly ionized in
2021.5 by the changing illumination from the hot component. 
The H$\alpha$ speckle imaging data of \cite{Liimets25}
from 2020.9, just taken half a year before our 2021.5 data,
detected clearly the cloud A$_{\rm SW}$ at about $r\approx 130$~mas
south of the binary and some weak extension around
$r \approx 180$~mas, which could be the clouds
A1$_{\rm SW}$ and A2$_{\rm SW}$ identified in Fig.~\ref{HaDustCent}d.
Unfortunately, the interpretation, whether
A1$_{\rm SW}$ and A2$_{\rm SW}$ are slow or fast moving clouds, remains unclear.

The cloud B$_{\rm SW}$ shows a slow, steady outward motion
of about $20~{\rm mas\,yr}^{-1}$ similar
to cloud C$_{\rm SW}$, which we describe later. 
The exact movement of B$_{\rm SW}$ is not easy to determine
because there are also some structural changes.
For epoch e$_1$ the cloud is well defined, but for the following two
epochs the cloud is elongated radially and fades, so that the flux
peak may not represent well
the bulk motion of this cloud. In addition, there appears in epoch e$_2$
a new component B1$_{\rm SW}$ which moves until e$_3$ like
component B$_{\rm SW}$ (Fig.~\ref{HaDustCent}). For the dust eclipse epoch
e$_4$ both components have disappeared most likely because they are
located at an azimuthal angle $\theta>225^\circ$ outside of the
radiation cone for the ionizing radiation from the hot component
(Sect.~\ref{SectHaObs}).

In the north there appears in epoch e$_3$ the transient cloud
A$'_{\rm N}$ at a separation of $r\approx 210$~mas. This cloud is
again practically absent for e$_4$ and just a weak trace is visible at
$r\approx 300$~mas indicating a relatively large
proper motion of $33~{\rm mas\,yr}^{-1}$.

We note that our map of epoch e$_3$ for 2018.8 matches very well
the ALMA 0.9~mm continuum map from 2017.9 presented in \cite{Bujarrabal18}. 
The transient cloud A$'_{\rm N}$ is in their map at a separation
of $\approx 180$~mas confirming its outward motion. The
agreement of the H$\alpha$ map
from 2018.8 is also in excellent with the JVLA 7~mm continuum map
of \citet{Cotton21} for 2019.6 about 9~months, except for the transient
cloud A$'_{\rm N}$, which was not detected.  

For epoch 2014.8 there is a narrow arc on the NW side just outside
of the central nebula extending from about $0.05''$ to $0.2''$
also present in Fig.~7 of Paper I. This feature is absent
in our H$\alpha$ map from 2016.6 but very interestingly it is detected
with HST in the Mg~{\sc ii} filter ($\lambda= 280$~nm) for the
epoch 2017.8 by \citet{Huang23} and they call this feature
``Hinge''. This indicates that gas structures may disappear in H$\alpha$
because the ionization degree changes, but they are still
present. This feature is reviewed in the next section in connection
with the dust polarization detected in this region.

\subsection{Scattering polarization near the binary}

The circumstellar polarization maps $Q_\phi^{\rm ex}$ for the central
$0.8''\times 0.6''$ region of R~Aqr
are plotted in Fig.~\ref{HaDustCent} (right panels) for the four
epochs. The $Q_\phi^{\rm ex}$ signal is particularly strong near the center,
because there the irradiation by the Mira and the density of the scattering
dust are high and then drop rapidly with
radial separation. This region shows a highly structured signal and strong
temporal variations. 

\subsubsection{Structural changes for the central $Q_\phi^{\rm ex}$ polarization}

For the first two epochs about 1.5 times more scattering polarization
is observed on the east side of the Mira
when compared to the west (Fig.~\ref{HaDustCent}). 
This ratio is 2.7 for epoch e$_3$ mainly because
of a the strong signal just
east of the Mira. For the obscuration event in 2021.5 there appeared
a high surface brightness region on the west extending 
out to $r\approx 0.3''$, and the $Q_\phi^{\rm ex}$ fluxes
on the east and west sides are practically equal.
The overall $Q_\phi^{\rm ex}$
signal appears to be stronger for this epoch because the map is
roughly scaled to the normalized polarization $Q_\phi^{\rm ex}/I_\star$,
which is enhanced because of the strong extinction of the
intensity of the Mira $I_\star$.
The strong changes between epoch e$_3$ and e$_4$ are apparently the
result of and increase of scattering on
the west side and a suppression of signal east of the Mira during the
occultation event.

Several discrete structures are visible despite and show temporal changes 
despite the strong radial flux gradient and the variations
of the pulsating Mira.
In the northeast there is a local $Q_\phi^{\rm ex}$ maximum (a$_{\rm NE}$)
moving from epoch e$_2$ to e$_3$ with a proper motion of roughly
10~mas\,yr$^{-1}$.
This extension has expanded further in epoch e$_3$, but there
is no local peak anymore, while in e$_4$ its surface brightness
dropped strongly, while a faint extension still expanded further
outward similar to a cloud that dispersed radially. The peak of the
cloud a$_{\rm W}$ on the west side shows also a steady radial motion
from epoch e$_2$ to e$_4$ with about 10~mas\,yr$^{-1}$. We later note
that these velocities
are smaller than for dust clouds at larger distance and suspect that this
could hint to an acceleration of the dust clouds by 
radiation pressure. 

An interesting dust cloud is a$_{\rm NW}$ located in the northwest
at a separation of about 100~mas. This cloud is too faint to
be seen in Fig.~\ref{HaDustCent}, but it is clearly visible in
the deeper map presented in Sect.~\ref{SectAn} (Fig.~\ref{FigDustHaNorth}).
This dust cloud can be associated with the 
nearby H$\alpha$ ``Hinge'' feature mentioned above. 
\cite{Huang23} detected at this position a strong Mg\,{\sc ii} emission
with HST and suggest that the presence of ionized
magnesium points to dust sputtering by shocks in the ionized outflow of R~Aqr.
This is supported by cloud a$_{\rm NW}$ which proofs the presence
of dust at this location. The ``Hinge'' emission visible in 2014.8
is offset by 46~mas from the
faint dust cloud toward the center (Fig.~\ref{HaDustCent}e), like a dusty
gas cloud partly ionized on the side facing the central binary.
In later epochs the H$\alpha$ emission disappeared but the dust
clouds a$_{\rm NW}$ is still present and moves radially outward.
More dust clouds with nearby H$\alpha$ emission exist around R~Aqr
and they are described later.

\paragraph{Comparison with CO maps.}
A lot can be learned from a comparison of the maps for the dust polarization
and the CO emission observed with ALMA. Overall there
is a very good agreement between the polarization maps shown
in Fig.~\ref{HaDustCent} panels (f) and (g) for 2016.6 and 2018.8
with the CO maps for the $J=3-2$ transition for 2017.9 presented
by \cite{Bujarrabal21} and for velocity channels close to the
systemic velocity for the CO $J=6-5$ transition for 2019.6
(Figs.~6 and B2 in \cite{GomezG24}).
The good correspondence indicates that the
CO line emission and the dust polarization features are
optically thin. We can estimate the radial velocity component
of the dust features from the CO maps based on the assumption that
the dust is well coupled to the large scale CO gas flows.
According to this, the bulk of the dust located about $0.2''$ to the west
is moving toward us with roughly
$\Delta {\rm RV}\approx -4~{\rm km\,s}^{-1}$,
while the dust within $0.1''$ on the east is receding by
$\approx +2~{\rm km\,s}^{-1}$ with respect to the systemic 
${\rm RV}_0=-24.9~{\rm km\,s}^{-1}$
\citep{Gromadzki09}. In addition, the appearance 
of the strong dust scattering polarization on the west side in epoch e$_4$
can be associated with an approaching, bright CO cloud at this position.
For this epoch the hot component is in front of the Mira and the
approaching CO cloud is trailing the hot star in the orbital plane
according to the proposed for the
ALMA CO velocity maps by \cite{Bujarrabal21}.

\paragraph{Comparison with AGB star simulations.}
The polarization structure within about $0.1''$ on the east side of the Mira
shows an expanding dust bubble in the east-southeast, 
which grows and expands during the 6.7~years from about
$r\approx 60$~mas to 100~mas or with $6~{\rm mas\,yr}^{-1}$. This is a bit
slower than the $10~{\rm mas\,yr}^{-1}$ for the transient northeast
bubble a$_{\rm NE}$ described above. With the adopted distance of
$d=218$~pc this translates into dust structures with a diameter
of about 10~AU, expanding with
a projected velocity of about 5 to $10~{\rm km\,s}^{-1}$ at a separation
of $r=10-30$~AU from the center of the Mira star. The morphology,
size and expansion velocities of these dust structures
are very similar to the model results for dust forming clouds in
AGB stars with a dust-driven wind as obtained by the 3d
radiation-hydrodynamical calculations of \cite{Freytag23}
shown in their Figs.~2 and 3. Of course, the comparison of the
bubbles in the polarimetric image just resolved by about five
resolution elements on a scale of about 20~AU with a high-resolution
slice through a model is difficult. According to the simulations the
expanding bubbles produce dense dust concentrations, and R~Aqr shows
indeed such features further out as described later.
The structures on the west side of the Mira are different but
this can be attributed to the presence of the companion and
the dust cloud comoving with the hot star discussed above. 

\begin{figure}
  \includegraphics[trim=2.0cm 16.5cm 9.cm 2.8cm,clip,width=8.5cm]{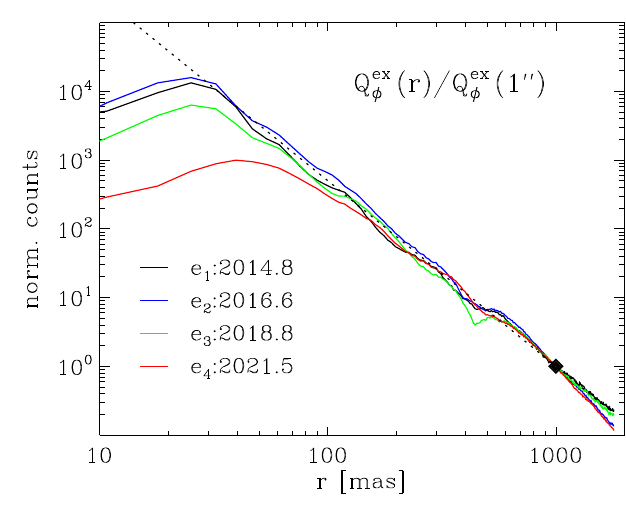}  
  \caption{Normalized radial profiles for the azimuthal polarization
    $Q_\phi^{\rm ex}(r)$ for all four epochs. The dotted line gives a
    power-law function of $\propto r^{-2.7}$. All curves have been normalized
    to $Q_\phi^{\rm ex}(1'')$ marked with a point.}
\label{FigRadQphi}
\end{figure}

\subsubsection{Polarimetric measurements}
The surface brightness of the circumstellar polarization
$Q_\phi^{\rm ex}(r)$ in R~Aqr decreases rapidly
by about a factor of 300 to 1000 from $r\approx 0.15''$ to $1.5''$
according to the contours in Fig.~\ref{HaDust14}.
This fall-off is illustrated in Fig.~\ref{FigRadQphi} for all four epochs
with radial profiles for the azimuthally averaged polarization
$Q_\phi^{\rm ex}(r)/Q_\phi^{\rm ex}$($r$=$1''$) normalized to the signal
at $1''$. For $r>0.1''$ all four profile show the same power law of
$Q_\phi^{\rm ex}(r)\propto r^{-2.7}$ indicated by the dotted line.
This is close to the expected power law of $Q_\phi^{\rm ex}(r)\propto r^{-3}$
for an optically thin, spherically symmetric outflow of homogeneous
dust particles with a density law of $\rho(r)\propto r^{-2}$. However,
a detailed interpretation would require modeling for the non-axisymmetric
dust distribution in R~Aqr. 

For $r\lapprox 30$~mas there is a turnover because
the circumstellar polarization is no more resolved
and suffers from substantial polarimetric cancellation
\citep{Schmid26}. There is a strong reduction of the polarization at
$r\lapprox 40$~mas for e$_4$ when compared
to the two first epochs, partly because of the dust obscuration event
and partly because of the reduced AO performance. A small flux
reduction is already for epoch e$_3$.

The derived azimuthally averaged $Q_\phi^{\rm ex}(r)$ profiles in
Fig.~\ref{FigRadQphi} depend hardly on the applied
zero-point offsets. The subtractions of 
$\langle q_0 \rangle\,I(\alpha,\delta)$
and $\langle u_0 \rangle\,I(\alpha,\delta)$ in 
Eqs.~\ref{EqQex} and \ref{EqUex} introduce symmetric
positive-negative quadrant patterns for $Q_\phi^{\rm ex}(\alpha,\delta)$
with zero net signal,
and this does not change significantly the azimuthally averaged
profiles $Q_\phi^{\rm ex}(r)$ as described in \cite{Schmid26}. 
Stokes $Q$ and $U$ offsets introduce a quadrant pattern in the
$Q_\phi^{\rm ex}(\alpha,\delta)$ map and change the azimuthal distribution
of the $Q_\phi$ signal but small scale clouds keep their polarization
structure, and one can still follow their evolution during the four
epoch \citep{Schmid26}.

\begin{table}
  \caption{ZP-corrected I-band polarimetry for R~Aqr.}  
\label{DataPolAp}
\begin{tabular}{lcccc}
\noalign{\hrule}
\noalign{\smallskip\hrule\smallskip}
parameter & 2014.8 & 2016.6 & 2018.8 & 2021.5 \\
\noalign{\smallskip\hrule\smallskip}
\noalign{\noindent total polarization for $r<1.5''$ \smallskip}
$\Sigma Q_\phi^{\rm ex}/I_\star$ [\%] 
                       & $+0.60$ & $+0.78$ & $+1.12$  & $+2.01$ \\
\noalign{\smallskip}
$\Sigma Q^{\rm ex}/I_\star$ [\%]& $+0.19$ &  $+0.36$ &  $+0.63$ & $+1.60$  \\
$\Sigma U^{\rm ex}/I_\star$ [\%] & $-0.12$ &  $+0.15$ &  $+0.20$ & $+1.03$ \\
${\cal{P}}^{\rm ex}/I_\star$ [\%]    & 0.22  & 0.39        & 0.66     & 1.90  \\
$\theta^{\rm ex}$        & $-16^\circ_{\pm 5^\circ}$ & $+11^\circ_{\pm 5^\circ}$
                               & $+9^\circ_{\pm 5^\circ}$ & $+16^\circ_{\pm 5^\circ}$ \\                       
\noalign{\smallskip}
${\cal{P}}^{\rm ex}/\Sigma Q_\phi^{\rm ex}$
                        & 0.37    & 0.50   & 0.59  & 0.95  \\
                        
\noalign{\smallskip}
\noalign{\noindent polarization for $r_{\rm in}<0.30''$ \smallskip}

$\Sigma_{\rm in} Q_\phi^{\rm ex}/I_\star$ [\%] &
                         $+0.53$ &  $+0.68$ & $+0.95$  & $+1.46$  \\
\noalign{\smallskip}
$\Sigma_{\rm in} Q^{\rm ex}/I_\star$ [\%]\hspace{-4mm}
                        & $+0.13$ &  $+0.23$ &  $+0.42$ & $+1.00$  \\
$\Sigma_{\rm in} U^{\rm ex}/I_\star$ [\%]\hspace{-4mm}
                        & $-0.07$ &  $+0.10$ &  $+0.15$ & $+0.66$  \\
${\cal{P}}^{\rm ex}_{\rm in}/I_\star$ [\%]
                        & $0.15$  & $0.25$ & $0.45$  & $1.19$     \\
$\theta^{\rm ex}_{\rm in}$  & $-14^\circ_{\pm 10^\circ}$ & $+5^\circ_{\pm 5^\circ}$
                       & $+10^\circ_{\pm 5^\circ}$ & $+16^\circ_{\pm 5^\circ}$ \\
\noalign{\smallskip}
$\Sigma_{\rm in}I/I_\star [\%]$ & 64   & 68   &  69    & 70  \\
\noalign{\smallskip}
\noalign{\noindent polarization for $0.30''< r_{\rm out} <1.50''$ \smallskip}
\noalign{\smallskip}
$\Sigma_{\rm out} Q_\phi^{\rm ex}/I_\star$ [\%] &
                         $+0.07$ &  $+0.10$ & $+0.17$  & $+0.54$  \\
\noalign{\smallskip}

$\Sigma_{\rm out} Q^{\rm ex}/I_\star$ [\%]\hspace{-4mm}
                        & $+0.06$ &  $+0.13$ &  $+0.21$ & $+0.59$  \\
$\Sigma_{\rm out} U^{\rm ex}/I_\star$ [\%]\hspace{-4mm}
                        & $-0.05$ &  $+0.05$ &  $+0.05$ & $+0.37$  \\
${\cal{P}}^{\rm ex}_{\rm out}/I_\star$ [\%]
                        & 0.08    &  0.14    &  0.22    & 0.69 \\
$\theta^{\rm ex}$
                       & $-19^\circ_{\pm 15^\circ}$ & $+11^\circ_{\pm 10^\circ}$
                               & $+7^\circ_{\pm 7^\circ}$ & $+15^\circ_{\pm 5^\circ}$\\
\noalign{\smallskip}
\noalign{\noindent values relative to
    $\Sigma_{\rm out} Q_\phi^{\rm ex}$ \smallskip}
$I_\star/\Sigma_{\rm out} Q^{\rm ex}$ 
         & $\approx 1400$ & $\approx 1000$ & $\approx 590$ & $\approx 190$ \\  
$\Sigma_{\rm in} Q^{\rm ex}/\Sigma_{\rm out} Q^{\rm ex}$ 
         & $\approx 7.5$ & $\approx 6.8$ & $\approx 5.6$ & $\approx 2.7$ \\  
\noalign{\smallskip}
\noalign{\noindent center polarization\tablefootmark{a} \smallskip}
$\langle q_0\rangle$ [\%] & $+0.11$ &  $-0.10$ &  $+0.50$ &  $-0.86$
                        \\
$\langle u_0\rangle$ [\%] & $-0.15$ &  $-0.36$ &  $-0.13$ &  $-1.00$
                         \\
$p_0$ [\%]             & 0.19      &  0.37   &   0.52    &  1.31  \\    
$\theta_0$             & $-29^\circ_{\pm 5^\circ}$ & $-53^\circ_{\pm 5^\circ}$
                       & $-7^\circ_{\pm 5^\circ}$ & $-65^\circ_{\pm 5^\circ}$ \\

\noalign{\smallskip\hrule\smallskip}

\end{tabular}

\tablefoot{ZP correction based on $\langle q_0 \rangle$ and
  $\langle u_0 \rangle$ derived from a few central pixels for the central
  star. Estimated measuring uncertainty for the Stokes parameters
  $\Sigma_iQ^{\rm ex}/I_\star$ and $\Sigma_iU^{\rm ex}/I_\star$ are about
  $\pm 0.03~\%$. 
  \tablefoottext{a}{The values have not been corrected for the residual telescope
    polarization of about $0.3~\%$.}}
\end{table}

The measured polarization for the circumstellar scattering is
given for the four epochs in Table~\ref{DataPolAp}.
The azimuthal polarization $Q_\phi^{\rm ex}$ for the resolved
circumstellar signal and the corresponding
Stokes polarization $Q^{\rm ex},U^{\rm ex}$,
corrected for the polarization of the central source, are derived for
a large circular aperture $\Sigma$ with $r=1.5''$, for a small
circular aperture $\Sigma_{\rm in}$ with $r=0.3''$ covering the
the inner region shown in Fig.~\ref{HaDustCent}, and for an annular
aperture $\Sigma_{\rm out}$ with $0.3''<r<1.5''$.
These values are normalized to the total systen intensity
$I_\star=\Sigma I$ within
an aperture with $r=1.5''$. Also given are
the fractional Stokes values for the central source $\langle q_0 \rangle$ and
$\langle u_0 \rangle$, which were used for the zp correction 
(Eqs.~\ref{EqQex} and \ref{EqUex}), and some
diagnostic parameters discussed below.

Most polarization parameters depend on the PSF of the observations
as described in \cite{Schmid26}. According to this study
and for a PSF representative for the epochs e$_1$, e$_2$, and e$_3$, the
measurable $\Sigma_{\rm in} Q_\phi^{\rm ex}$ signal for the inner
region $0.025''<r<0.3''$ is about a factor of two to three smaller
than the intrinsic circumstellar polarization because of the
limited spatial resolution. 
The signal in the ``outer'' annulus $\Sigma_{\rm out} Q_\phi^{\rm ex}$
underestimates the intrinsic value by about 10~\% to 30~\%.
For epoch e$_4$ an even stronger polarimetric degradation $Q_\phi^{\rm ex}$
must be expected because of the reduced AO performance,
particularly for the inner region $\Sigma_{\rm in}$.  

The integrated Stokes parameters including the signal of the central source
of $\Sigma Q^{\rm ex}/I_\star+\langle q_0\rangle\approx +0.8~\%$ and
$\Sigma U^{\rm ex}/I_\star+\langle u_0\rangle\approx 0.0~\%$ are practically not
affected by the PSF convolution, and the measured values represent well
the intrinsic values apart from the $\pm 0.3~\%$ uncertainty from the
undefined residual telescope polarization. However, this signal
is redistributed between the central source, the inner small aperture,
and the outer annulus.

Extensive model simulations would be required to retrieve the instrinsic
polarization of the R~Aqr system, but this is beyond the scope of this paper.
Nonetheless, we can still estimate well the relative polarimetric
changes between the first three epochs taken with similar PSFs, and
constrain the changes for the last epoch considering the reduced
PSF quality.

The normalized azimuthal polarization $\Sigma Q^{\rm ex}_\phi/\Sigma I_\star$
increases steadily from epoch e$_1$ to e$_3$ from 0.60~\% to 1.12~\% or
by almost a factor of two. The increase is about a factor of three
for epoch e$_4$ with respect to epochs e$_1$ and e$_2$
(Table~\ref{DataPolAp}).
If one assumes that the extra dust extinction in 2021 only
reduces the $I_\star$ intensity by $\Delta_{\rm ext}I_{\rm vis}=1.8^m$, but not the
circumstellar polarization $\Sigma Q^{\rm ex}_\phi$, then the ratio
$\Sigma Q^{\rm ex}_\phi/\Sigma I_\star$ should increase by a factor of about five.
Indeed, an increase of about a factor of six between the first two
epochs ${\rm e}_1,{\rm e}_2$
and the dust obscuration epoch e$_4$ is obtained for
$\Sigma_{\rm out} Q^{\rm ex}_\phi/\Sigma I_\star$ 
measured in the annulus from $r=0.3''$ to $1.5''$.
We therefore assumed, as in Fig.~\ref{FigRadQphi}, that the circumstellar
scattering signal $\Sigma_{\rm out} Q^{\rm ex}_\phi$ at larger separation
can be used as rough flux reference. The ratio
$\Sigma_{\rm out} Q^{\rm ex}_\phi/I_\star$
for epoch e$_3$ is a factor two higher than for epochs 
${\rm e}_1,{\rm e}_2$ indicating already for epoch e$_3$ a
substantial extinction of the central source.

The comparison between the polarization from the inner and
the outer scattering region yields a ratio of
$\Sigma_{\rm in} Q^{\rm ex}_\phi/\Sigma_{\rm out} Q^{\rm ex}_\phi \approx 7$
as average for epochs e$_1$ and e$_2$. This changes to $\approx 2.7$
for the eclipse epoch e$_4$. We estimate that about half of this
reduction can be attributed to the extinction of radiation from
the inner scattering region by the dust eclipse event, and about
half is lost because of the stronger polarimetric cancellation
induced by the lower PSF quality for epoch e$_4$. The observed
turnover of the $Q^{\rm ex}_\phi(r)$ profile in Fig.~\ref{FigRadQphi}
around $r\approx 40$~mas indicates that the dust cloud responsible for the
extinction event in R~Aqr has a diameter of up to 20~AU (80~mas).

Also the normalized Stokes parameters for the extended emission
$\Sigma Q^{\rm ex}/\Sigma I_\star,\,\Sigma U^{\rm ex}/\Sigma I_\star$
show an increase during the eclipse because of the
reduction of the stellar flux. For the large aperture $\Sigma$, one can derive
the net linear polarization flux and position
angle of the extended emission with
\begin{equation}
  {\cal{P}}^{\rm ex}= \bigl( (\Sigma Q^{\rm ex})^2 +(\Sigma U^{\rm ex})^2 \bigr)^{1/2},
  \,\, \theta^{\rm ex} = 0.5\, {\rm atan}(\Sigma U^{\rm ex}/\Sigma Q^{\rm ex})\,. 
\end{equation}
The ratio ${\cal{P}}^{\rm ex}/\Sigma Q_\phi^{\rm ex}$ can be used as
measure for the spatial distribution of the scattering polarization.
An axisymmetric geometry produces no net Stokes signal
and the ratio is ${\cal{P}}^{\rm ex}/\Sigma Q_\phi^{\rm ex}\approx 0$, while
the ratio approaches ${\cal{P}}^{\rm ex}/\Sigma Q_\phi^{\rm ex}\rightarrow 1$
for highly asymmetric circumstellar scattering regions, like an optically
thin edge-on disk \citep[e.g.,][]{Schmid21}. The Stokes polarization for the
extended scattering region $\Sigma Q^{\rm ex},\,\Sigma U^{\rm ex}$
and the corresponding azimuthal polarization $\Sigma Q_\phi^{\rm ex}$
depend according to Eq.~\ref{QphiEx} in a similar way on the spatial
resolution. Therefore we can use the ratio ${\cal{P}}^{\rm ex}/\Sigma Q_\phi$
to compare relative changes in the extended scattering geometry
from observations taken with similar but not necessary equal
spatial resolution.

Our data show a change from a low ratio
${\cal{P}}^{\rm ex}/\Sigma Q_\phi^{\rm ex}\approx 0.37$ for e$_1$ to a high
ratio of $\approx 0.95$ for the e$_4$ eclipse epoch or from
a more axisymmetric geometry to a dust distribution strongly
aligned with directions of about $+106^\circ$ and $+286^\circ$. This
corresponds roughly to the perpendicular direction with
respect to the measured polarization position angle of
$\theta^{\rm ex}=16^\circ$. This also confirms
the appearance of the scattering geometry in
Fig.~\ref{HaDustCent}(h) taken during the
eclipse event. This evolution could partly
be caused by the extinction of the more axisymmetric signal
with low ${\cal{P}}^{\rm ex}/\Sigma Q_\phi^{\rm ex}$ for the dust
formation region near the star
and partly by enhanced scattering polarization from dust near
the orbital plane with a value close to one for
${\cal{P}}^{\rm ex}/\Sigma Q_\phi^{\rm ex}$ becoming bright for this epoch
in agreement with the CO cloud trailing the hot star according to
the model of \cite{Bujarrabal21}.

Not much can be said about scattering properties of the dust
from the presented polarimetry. Accurate multiwavelength polarimetry
including corrections for instrumental effects would be required
for a meaningful characterization of the dust. This is very demanding
\citep[e.g.,][]{Tschudi21,Ma24} and beyond the scope of this work.
However, the dust in R~Aqr was modeled in detail based on
high quality infrared spectra \citep{Omelian20,Omelian25,Sankrit22}.
It is found that the dust in R~Aqr is quite typical 
for an oxigen-rich Mira star, but requires also grains with sizes up
to $2~\mu$m, which might be formed in the dense equatorial disk. 

\section{The outflow region}
\label{SectOutflow}

The circumstellar gas at separations
$r\gapprox 0.3''$ ($\gapprox 65$~AU) is dominated by 
a steady radial motion and we call this the outflow regions.
The VLT data cover this region only up to $1.5''-2''$
and the images show a strong radial decrease
for the polarization and H$\alpha$ brightness
(e.g., Fig.~\ref{HaDust14} and Paper~I) but also many
interesting structures as described in this section.

\begin{figure}
  \includegraphics[trim=2.0cm 12.8cm 3.0cm 3.5cm,clip,width=8.5cm]{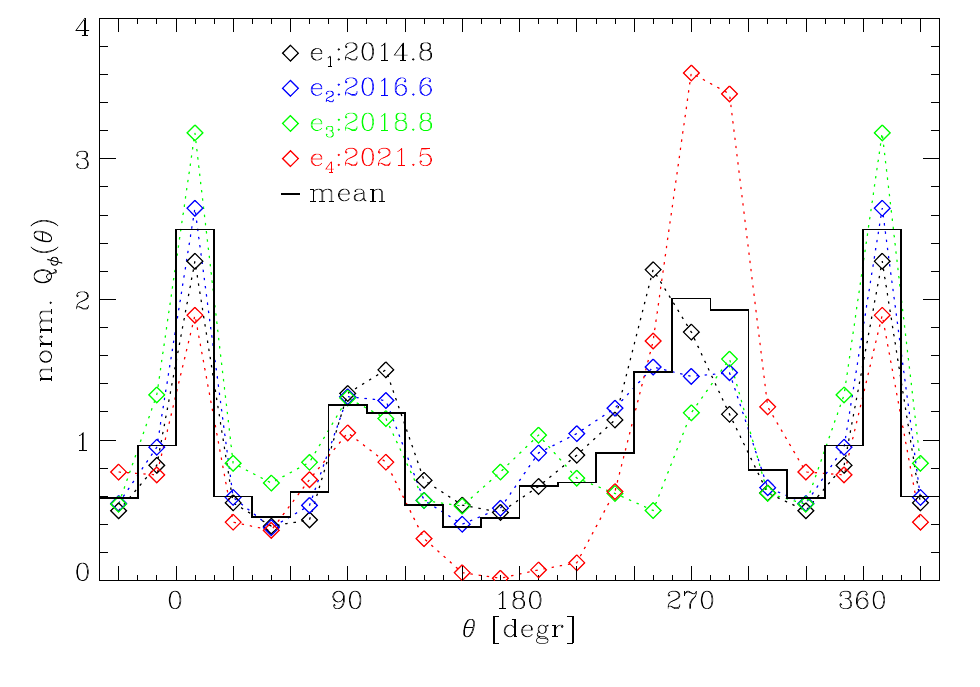}  
  \caption{Azimuthal distribution of the $Q_\phi^{\rm ex}(\theta)$ polarization
    for $300~{\rm mas} \leq r \leq 1500$~mas integrated in bins of
    $\Delta\theta=20^\circ$ for the four VLT
    epochs. The bin values have been normalized by $\Sigma_{\rm out}Q_\phi^{\rm ex}$
    so that the average value is one for each curve.
    Also given is the mean curve for all epochs.}
\label{Qphitheta}
\end{figure}

\subsection{Dust in the outflow region}

\begin{figure*}
  \includegraphics[trim=0.cm 0.8cm 1.5cm 1.5cm,clip,width=18cm]{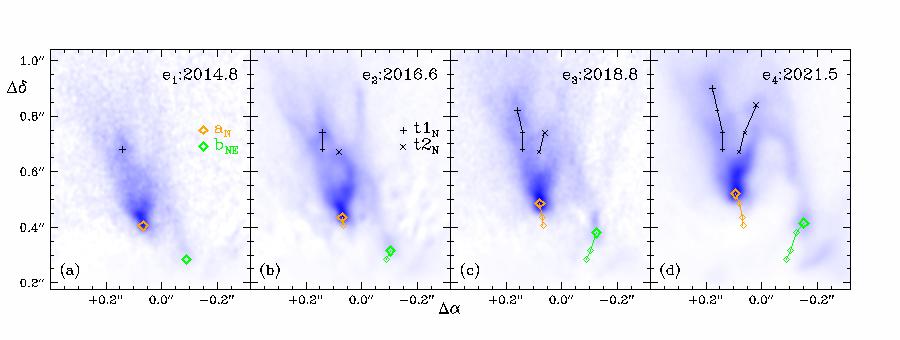}  
  \caption{Polarization maps $Q_\phi^{\rm ex}$ for the northern dust cloud a$_{\rm N}$
    with the two tail features t1$_{\rm N}$, t2$_{\rm N}$ for the four VLT epochs.
    The color is scaled with a south-north gradient to also show faint tail features.
    The flux is better represented in the contour maps in Fig.~\ref{FigDustHaNorth}. Features
    are indicated and replotted in later epochs with smaller symbols and lines to
    show motions.}
\label{DustNorth}
\end{figure*}

\subsubsection{The large scale polarization signal}

The radial profiles for the 
polarization are given by a steep power law of $Q_\phi^{\rm ex}(r)\propto r^{-2.7}$ 
(Fig.~\ref{FigRadQphi}), and therefore the
measured polarization for the outflow region $\Sigma_{\rm out} Q_\phi^{\rm ex}$
is quite small, only about
$\approx 0.12~\Sigma Q_\phi^{\rm ex}$, for epochs e$_1$ and e$_2$
(Table~\ref{DataPolAp}).
This even overestimates the intrinsic ratio
$\Sigma_{\rm out} Q_\phi^{\rm ex}/\Sigma Q_\phi^{\rm ex}$
because of the strong polarimetric cancellation 
near the central binary and values $<0.05$ should be expected
for e$_1$ and e$_2$ if the system is fully resolved. Of course,
the relative value $\Sigma_{\rm out} Q_\phi^{\rm ex}/\Sigma Q_\phi^{\rm ex}$
is higher for epoch e$_4$, because the dust eclipse event suppresses
strongly the intrinsic polarization from the bright, central
scattering region. 

The measured azimuthal distribution $Q_\phi^{\rm ex}(\theta)$ for the four
epochs is shown in Fig.~\ref{Qphitheta} for the integrated signals in
18 position angle wedges covering the outflow region
$300~{\rm mas}\leq r \leq 1500~{\rm mas}$. Each curve is normalized
so that the average value per bin is equal to one.
It must be noted that the derived $Q_\phi^{\rm ex}(\theta)$ curves
are based on the applied zp-corrected for the unresolved central object.
If this assumption is wrong, then $Q_\phi^{\rm ex}(\theta)$ is affected
\citep{Schmid26}. In addition, the PSF smearing of the polarization from the
bright inner integration region $\Sigma_{\rm in}$ can also
contaminate $Q_\phi^{\rm ex}(\theta)$ of the faint outer region.

We first consider the $Q_\phi^{\rm ex}(\theta)$ curves for the first
three epochs and assume that the convolution and calibration effects are small.
The curves show three maxima, a narrow strong peak at $\theta=10^\circ$
for the prominent northern dust cloud.
The other two peaks are broader and represent the flattened
overall dust distribution oriented in east-west direction with
maxima around $\approx 100^\circ$ and $270^\circ$. These maxima are about
twice the value of the minimum in the south around $180^\circ$.

The curve for the eclipse epoch e$_4$ shows a much enhanced
signal on the west side around $280^\circ$, a lower signal for
the south, and also a relative weak signal for the north cloud.
The increase in the west is caused by the strongly enhanced polarization
signal at $r<0.3''$ on the west side of the binary
(Fig.~\ref{HaDustCent}h). This cloud produces a strong Stokes $Q$ component,
and the PSF smearing therefore contaminates the
weak signal in the outflow region. This additional $Q$ signal
adds a positive contribution to $Q_\phi^{\rm ex}$ in the west, and negative
contributions to the south and north region \citep[see][]{Schmid26}.
Despite these convolution effects the general shape of the
$Q_\phi^{\rm ex}(\theta)$ curve for epoch e$_4$ is still qualitatively valid,
indicates more scattering polarization
in the east and west because of a flattened dust outflow, and the presence
of a prominent dust cloud in the north.

The flattened dust distribution in the
outflow region is in line with the detected 70~$\mu$m
east-west emission features from cold dust ($\approx 35$~K) of R~Aqr
seen on much larger scales of $r\approx 40''$ with the Herschel satellite
\citep{Mayer13}. This indicates that the
flattened dust outflow in R~Aqr prevailed for centuries. 

\begin{figure}
  \includegraphics[trim=0.cm 0.6cm 1.5cm 1.5cm,clip,width=8.5cm]{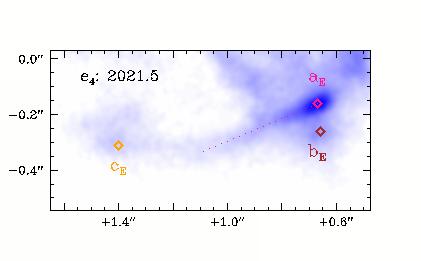}  
  \caption{Dust cloud a$_{\rm E}$ for epoch e$_4$ in the neutral Mira wind region
    on the east with the comet-like tail marked
    by a dotted line. The color is scaled
    with a west-east gradient to show the faint tail feature. 
    The additional dust clouds b$_{\rm E}$ and c$_{\rm E}$ are marked.
  }
\label{DustEastComet}
\end{figure}

\subsubsection{The comet-like dust clouds in the north}
\label{SectCometDust}
The most prominent dust feature in the outflow region is the
comet-like cloud a$_{\rm N}$
about 450~mas ($\approx 100$~AU) north of the binary
plotted in Fig.~\ref{DustNorth} for the four epochs. In this
figure the color table is scaled to the maximum of the bright
cloud core and a south-north gradient is applied to
illustrate the core and the weak tail features together.
The cloud a$_{\rm N}$ is located in the ionized outflow region
and there is also H$\alpha$ emission at this location. Later
in Sect.~\ref{SectAn} the a$_{\rm N}$ dust structure is reviewed
together with the H$\alpha$ emission.

The dust core a$_{\rm N}$ shows a steady radial motion of
$17.5~{\rm mas}\,{\rm yr}^{-1}$, while the 
two weak dust knots t1$_{\rm N}$ and t2$_{\rm N}$ in the tail
move about twice as fast (Fig.~\ref{DustNorth}).
The positions of t1$_{\rm N}$ and t2$_{\rm N}$ are difficult to determine
accurately, but the presence of higher velocity dust
is also supported by the motion of other faint tail features.

A backward extrapolation of the motion of these features,
assuming constant velocities, gives for the core a$_{\rm N}$ and the tail
knots around year 2002 the same position at $r\approx 0.2''$
north of the binary. This can be considered as point when the tail knots
t1$_{\rm N}$ and t2$_{\rm N}$ were released from the core. Extrapolating
further gives around year 1990 for a$_{\rm N}$ a position near the
binary where the dust cloud was probably formed. These are only
very rough indications about the life time of this comet like dust cloud. 

It is likely that the dust tail is optically thin and that dust particles
are charged because of the irradiation with ionizing
radiation from the hot component.
Very intriguing is the filamentary fan structure
of the tail which could be caused by the motion of charged
dust particles within a magnetized, outflowing plasma.
This suspicion is also supported by the dust cloud 
a$_{\rm E}$, which shows a narrow and straight dust tail
(Fig.~\ref{DustEastComet}).
This cloud is located on the east of the Mira in a region 
completely shielded from the radiation and the ionized outflow
of the hot component. This might explain the large
differences in the tail structures between these two comet-like clouds.

Another interesting structure in Fig.~\ref{DustNorth} is the dust
cloud b$_{\rm NW}$ and a dust filament just north of it.
Also b$_{\rm NW}$ moves radially away from the central binary
with a similar velocity as a$_{\rm N}$. The filament moves together
with b$_{\rm NW}$, but shows also a motion toward the west
and it seems that the filament is pushed by a
higher pressure bubble forming from 2014 to 2021 between
a$_{\rm N}$ and the filament. Apparently, this bubble expansion 
accumulates more scattering dust in the narrow filament as can be
inferred from the increased brightness of this feature in later epochs.

\begin{figure}
  \includegraphics[trim=0.3cm 0.8cm 1.7cm 1.5cm,clip,width=8.5cm]{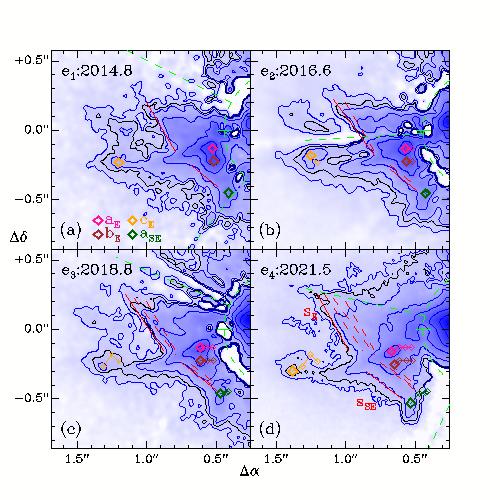}  
  \caption{Polarization map $Q_\phi^{\rm ex}$ on the east of R~Aqr for the four VLT epochs. 
    Blue contours give relative flux levels of $c\cdot 2^n$ with
    $n=0,1,2,...$, while one black contour per frame has been added to show 
    some structures better. The $Q_\phi^{\rm ex}$ clouds are identified
    with small symbols and lines
    to indicate cloud motions in later epochs. The dust density steps s$_{\rm E}$ and s$_{\rm SE}$ are
    marked. The dashed green lines indicate regions affected by PSF
    features.  }
\label{DustEast}
\end{figure}

\subsubsection{Dust cloud structures in the east}
\label{SectDustEast}
Figure~\ref{DustEastComet} shows for epoch e$_4$
a small region on the east side of the binary illustrating the rich
structure in the dust distribution.
A more comprehensive view is given by the four $Q_\phi^{\rm ex}$ maps
in Fig.~\ref{DustEast} showing the local
$Q_\phi^{\rm ex}$ flux maxima for the clouds a$_{\rm E}$, b$_{\rm E}$, c$_{\rm E}$
and a$_{\rm SE}$, and two
straight $Q_\phi^{\rm ex}$-intensity steps s$_{\rm E}$ and s$_{\rm SE}$.
The cloud a$_{\rm E}$ is about four times fainter than the
cloud core a$_{\rm N}$ in the North, while b$_{\rm E}$, c$_{\rm E}$ and
a$_{\rm SE}$ are about 15 times fainter.

All feature move radially away with a proper motion of about
$20~{\rm mas\,yr}^{-1}$ from the Mira, which is used as
$(0,0)$ reference point. Because
the Mira moves on its orbit to the west the
eastward motions of the dust features are about 15~\% smaller when
considering their motion relative to the center of mass instead
of the motion relative to the Mira. The analysis of the dust cloud motions
will be presented in Sect.~\ref{SectPMdust}. The locations of
the weaker clouds on the east are not so well defined because of
noise effects. Despite this, their proper motion can be derived
quite accurately because of the long temporal base line. 

The outward moving dust density steps s$_{\rm E}$
and s$_{\rm SE}$ look like a circular arc in the meridional
density profile obtained for the edge-on view of
a spiral pattern in simulations for
AGB binary stars with parameters similar to R~Aqr
\citep[e.g.,][]{KimH24,KimH12,Mastrodemos99}. The density steps
are predicted to move roughly with the sum of the velocities for
the stellar wind and the orbital motion of the mass losing
star $v_{\rm wind}+v_{\rm orbit}$, and steps are radially separated
by about $(v_{\rm wind}+v_{\rm orbit})\cdot P$, where $P$ is the orbital period
\citep{KimH12}. In R~Aqr the wind velocity dominates
($v_{\rm wind}\gg v_{\rm orbit}$), and this would give steps separated by
about $20~{\rm mas\,yr}^{-1} \cdot 44~{\rm yr}\approx 0.9''$.
This is about the separation of the step from the central binary, possibly
indicating that they where formed about one orbital period ago. 
This estimate should be considered cautiously, because of the
high eccentricity of the R~Aqr orbit and this is not considered in
such models.

\subsubsection{Dust structures in the west}
\begin{figure}
  \includegraphics[trim=0.cm 0.6cm 1.5cm 1.5cm,clip,width=8.5cm]{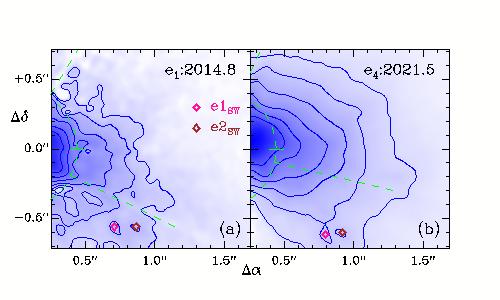}  
  \caption{
    Polarization maps $Q_\phi^{\rm ex}$ for the west of R~Aqr for the first and last VLT epochs.
    Flux contours for levels $c\cdot 2^n$ with
    $n=0,1,2,...$ are given, and two clouds are indicated. 
    The dashed green lines mark regions
    affected by PSF features.
  }
\label{DustWest}
\end{figure}

The scattering polarization at $r\gapprox 0.3''$ west of
the central binary is shown in Fig.~\ref{DustWest} for the
first and last observing epochs. The $Q_\phi^{\rm ex}$ signal
is dominated by a rather smooth radial gradient
with much less structures when compared to the
east side.
There might be dust clouds present,
but then they are not strong enough to produce a local maximum
on top of the rather steep $Q_\phi^{\rm ex}$ gradient. Only in the
south-west near the boundary of the plotted area in Fig.~\ref{DustWest},
there are two local $Q_\phi^{\rm ex}$ maxima from the clouds e1$_{\rm SW}$
and e2$_{\rm SW}$. They are described later together with
the H$\alpha$ emissions seen in this region.

\subsection{The H$\alpha$ outflow}

\begin{figure*}
\includegraphics[trim=0.cm 0.3cm 0.cm 1.5cm,clip,width=18.8cm]{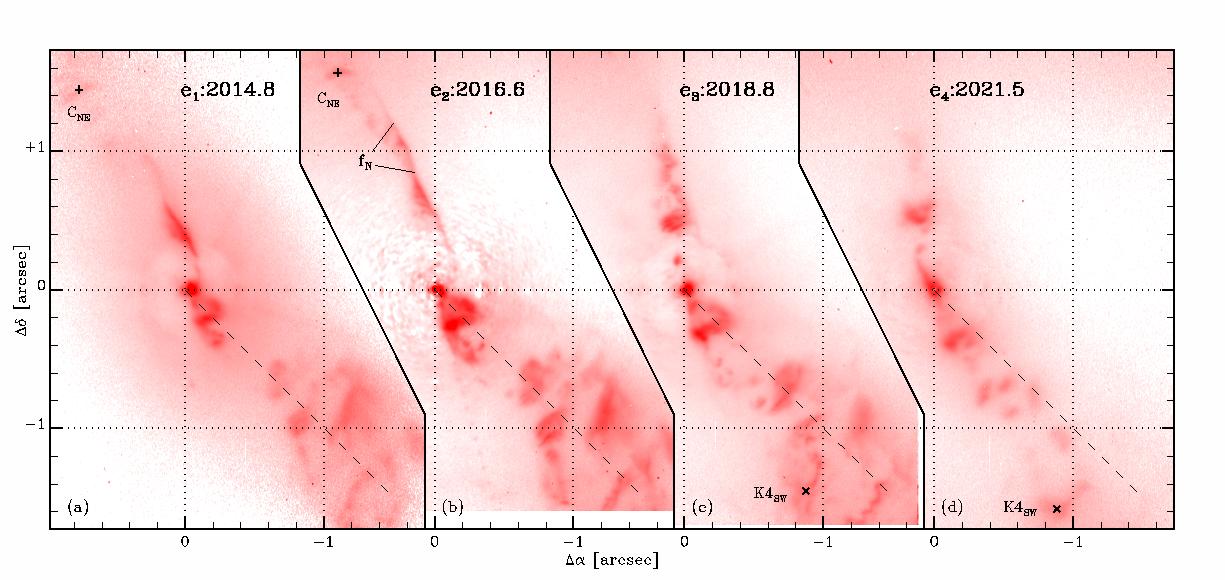}  
\caption{Differential $I_{{\rm H}\alpha}$ flux of R~Aqr for the four VLT epochs.
  The color is scaled with separation using power law functions
  $\propto r^\alpha$ with $\alpha\approx 2$
  to show the bright inner and faint outer cloud
  together. The dashed line gives the SW direction $\theta=225^\circ$. The
  two clouds C$_{\rm NE}$ and K4$_{\rm SW}$ 
  are marked to facilitate comparison with the HST maps.
  }
\label{FourFigHaVLT}
\end{figure*}

\subsubsection{Previous data for the ionized outflow}
We review briefly the literature on the collimated
outflow of R~Aqr since its detection around 1980 \citep{Herbig80,Sopka82},
to put the new H$\alpha$ data presented in this work into context.
The outflow extends over more than 30~arcsec
from the binary in NE and SW directions and seems to feed the
filamentary structures of the extended northern and southern emission
nebula.
There exists a descriptive model for the large scale nebular
structure consisting of the curved outflow from
the central binary and two extended nebular
bubbles, one north and one south, which are also called the hourglass
structure
\citep[e.g.,][]{Solf85,Liimets18,Santamaria24}.
The nebular features outside a few arcsec are expanding 
with about $150~{\rm mas\,yr}^{-1}$,
but the motion for some filaments and clouds deviates from a strictly radial
expansion \citep{Hollis97a,Liimets18}.
The kinematic age of the extended structures is a few hundred years
\citep[e.g.,][]{Santamaria24}.

The ionized outflow has been observed on subarcsecond
scales with radio interferometry of the free-free emission
\citep{Kafatos89,Dougherty95,Maekinen04,Cotton21} and high-resolution
imaging of UV and visual emission
lines, mostly with HST \citep{Paresce94,Hollis97a,Melnikov18,Huang23}.
In the years 1980 to about 2005 the NE outflow dominated the emission
within two arcsec of the central binary
\citep[e.g.,][]{Hollis85,Maekinen04,Kellogg07}
while after 2012 the SW region is clearly
dominant (Paper I). The
cloud emissions at $r\lapprox 2''$
vary on timescales of years and if the same structures
could be recognized for different epochs then the derived proper
motion was smaller than about $50~{\rm mas\,yr}^{-1}$
\citep{Paresce94,Dougherty95,Maekinen04,Melnikov18}.
The properties of the H$\alpha$ outflow at $r<2''$ can now
be investigated in much more detail thanks to the better temporal coverage
provided by the VLT and HST data.
The observations from 2014.8 were already described in
Paper I and we use the cloud identifications from that work and extend
this as needed.

\subsubsection{H$\alpha$ observations from 2014 to 2023}
\label{SectHaObs}
The differential $I_{{\rm H}\alpha}$
maps in Fig.~\ref{FourFigHaVLT} for the four epochs are plotted
with a radial scaling of the color to illustrate the structures for the
bright inner and faint outer features together.
The SW outflow shows for the first three epochs 
many clouds near the dashed SW-line ($\theta=225^\circ$)
and several clouds can be recognized in consecutive epochs because their
structures remain similar, and they show a steady motion in SW direction.
More details and some additional
faint, small clouds are visible for the data from
epoch e$_2$ and e$_3$ taken under better observing conditions.

There happened a dramatic change for epoch e$_4$ during the
dust eclipse as all clouds west of position angle
$\theta=225^\circ$ have practically disappeared, and
the ionized outflow has shifted significantly toward the south.
On average the clouds for e$_4$ are located around $\theta \approx 210^\circ$
and already in e$_3$ first signs of this change can be recognized
with the appearance of new clouds east of $\theta=225^\circ$ and
an overall brightness decrease on the west side. 

The NE outflow shows less H$\alpha$ emission as well as fewer
structural changes with time. In particular the bright
H$\alpha$ cloud A$_{\rm N}$, which is associated with the
prominent dust concentration a$_{\rm N}$, shows an interesting
evolution as described later in Sect.~\ref{SectAn}.

North of A$_{\rm N}$ is an elongated structure, which is 
rather faint for epoch e$_1$. For e$_2$
it shows a narrow, continuous filamentary structure marked as f$_{\rm N}$
with a length $>1''$ and width $<0.05''$ reaching up to
the cloud C$_{\rm NE}$. This is a unique feature in our data and we suspect
a link with a magnetic field in the ionized outflow.
Two years later, for e$_3$ the bright parts of the filament turned into
a knotty structure, perhaps because some of the ionized gas along
the filament recombined. According to the HST data described 
below the cloud C$_{\rm NE}$ moved further to the north.
We just point out this special, possibly magneto-hydrodynamic
feature f$_{\rm N}$ in the inner outflow region, which could
also be related to the fan like tail structures of the dust cloud
a$_{\rm N}$ described in Sect.~\ref{SectCometDust}.
There are more such filamentary structures in the
extended nebula of R~Aqr as shown in the deep H$\alpha$ and
[N~II] images by \cite{Melnikov18} and \cite{Liimets18} but a further
investigation of possible magnetic structures is beyond the scope
of this work.

\begin{figure*}
\includegraphics[trim=0.cm 0.3cm 0.cm 1.5cm,clip,width=18.8cm]{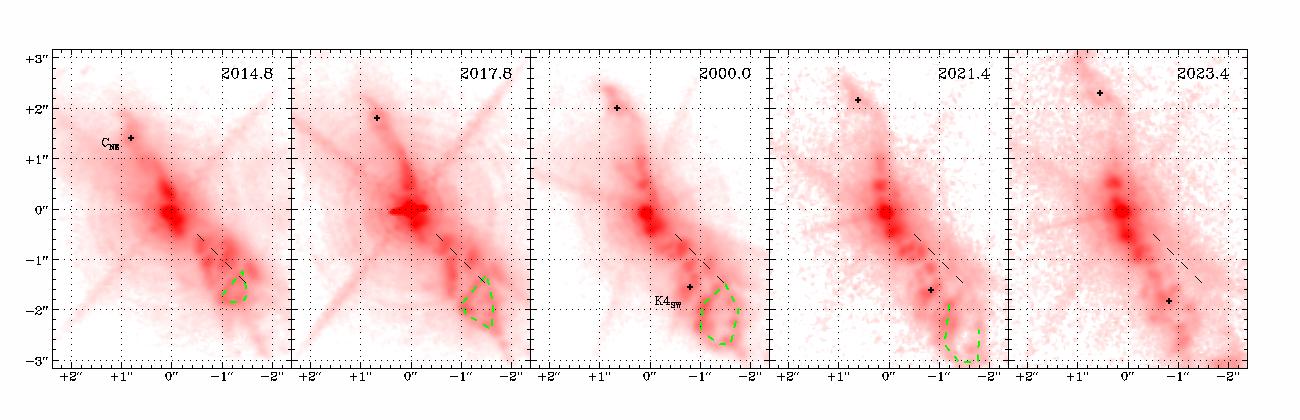}  
\caption{HST flux maps of R~Aqr in the F656N filter for five HST epochs.  
  The color scale is logarithmic.
  The dashed line gives the SW direction $\theta=225^\circ$, and the
  clouds C$_{\rm NE}$ and K4$_{\rm SW}$ are marked to facilitate comparison with the VLT maps.
  Also indicated, with a dashed green line, is the bubble H$_{\rm SW}$+I$_{\rm SW}$+J$_{\rm SW}$
  used by \cite{Huang23} for a cloud motion measurement.} 
\label{FiveFigHST}
\end{figure*}

\paragraph{H$\alpha$ observations from HST.}
The R~Aqr outflow was also observed for 5 epochs with
HST and the H$\alpha$ maps obtained for the central $4'' \times 6''$
are plotted in Fig.~\ref{FiveFigHST}. They are ideal
to check the results from SPHERE/ZIMPOL and complement the
temporal coverage. The plotted HST images were
all taken with the F656N line filter including the continuum
of the Mira star. We did not attempt to subtract the continuum but
note that all HST observation were taken when the Mira was
close to its brightness minimum (Fig.~\ref{lightcurve}).
The spatial resolution of HST is
lower than for the VLT, but the data are more sensitive for
the detection of faint features
and the photometric fidelity is higher.
In addition, there exist also deeper HST data for the same
epochs taken in many filter containing extensive additional
information about R~Aqr as described in \citet{Melnikov18} and
\citet{Huang23}.

We find an excellent agreement between the H$\alpha$ cloud maps
of HST and VLT as already described in Paper I.
The HST data confirm the angular shift of the
average position angle for the emitting H$\alpha$ clouds in the SW
jets from 2017 to 2021, and they show that this trend continues even
further for the 2023.6 epoch toward an average angle of
about $\theta\approx 205^\circ$
\citep[see also][]{Lucas24}.
The selected field of view for the HST maps include
the cloud C$_{\rm NE}$ in the north
and the bubble H$_{\rm SW}$+I$_{\rm SW}$+J$_{\rm SW}$ in the southwest
(Fig.~\ref{FiveFigHST}) because we use these features to refine the
previously described dependence of cloud motions
with distance from the binary.

In the northern outflow the H$\alpha$ cloud A$_{\rm N}$ for epoch
2014.8 can be recognized and thanks to the higher sensitivity
also the 
elongated extension out to cloud C$_{\rm NE}$ is visible.
From 2014 to 2023 this elongated structure moves toward the west
and expands in north
direction. It seems that this arc is pushed in NW direction by
an unseen gas structure located about $1.5''$ to the NE of R~Aqr. 

\subsection{Detailed descriptions of H$\alpha$ cloud structures}

The VLT data of R~Aqr provide
$I_{{\rm H}\alpha}$ and $Q_\phi^{\rm ex}$
maps of dusty plasma with unprecedented resolution of about 5~AU,   
thus enabling different kinds of structures to be distinguished
and a more detailed description. Quantitative
results for the cloud distribution and motions are given in
Sect.~\ref{SectPMall}. 

\subsubsection{The H$\alpha$ emission near the dust cloud a$_{\rm N}$}
\label{SectAn}

\begin{figure*}
\includegraphics[trim=0.cm 0.3cm 0.cm 1.5cm,clip,width=18.8cm]{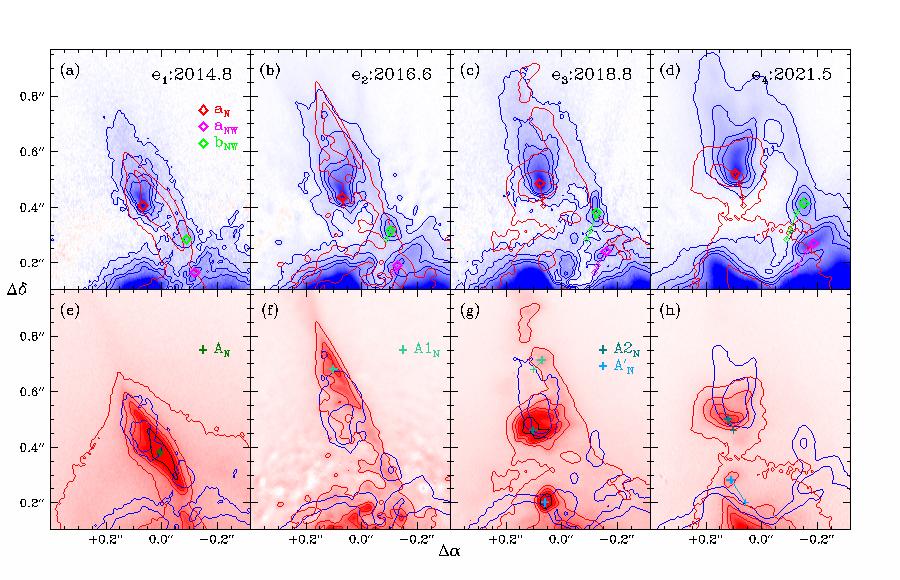}  
\caption{ I-band
  polarization $Q_\phi^{\rm ex}$ (top row) and differential $I_{{\rm H}\alpha}$ flux (left) 
  of the cloud in the north of R~Aqr for the four VLT epochs. 
  Relative flux contours for $Q_\phi^{\rm ex}$ and $I_{{\rm H}\alpha}$ are given
  for levels $c\cdot n$ with $n=1,2,3,4$. In panels (e) and (g), an
  extra contour for $n=6$ is given in black.
  For comparison some contours for the polarization are repeated
  in the H$\alpha$ maps, and vice versa.
  $Q_\phi^{\rm ex}$ and H$\alpha$ clouds are identified and replotted in later epochs
  with small symbols and lines to show cloud motions.} 
\label{FigDustHaNorth}
\end{figure*}

The H$\alpha$ emission and the dust scattering polarization
located about $0.5''$ north of the binary
are outstanding as bright clouds in the $I_{{\rm H}\alpha}$ and $Q_\phi^{\rm ex}$
maps, respectively, and contour maps are plotted in
Fig.~\ref{FigDustHaNorth}.

For the first three epochs there is an elongated H$\alpha$ structure on the
west side of the dust polarization peak a$_{\rm N}$.
The bright emission peak A$_{\rm N}$ is located for epoch e$_1$ about 70~mas 
on the SW side of the scattering dust cloud. This elongated
emission is still present for e$_2$
but weaker, and the new peak location A1$_{\rm N}$ is about 240~mas north
of the dust cloud. This structure is still present for e$_3$
and has disappeared in the VLT map for e$_4$, but it
is still visible as faint feature in the HST data from that year
(Fig.~\ref{FiveFigHST}d).

For the epoch e$_3$, there appeared the new, bright H$\alpha$ emission
A2$_{\rm N}$ on the SE side of the dust cloud, and the $I_{{\rm H}\alpha}$
and $Q_\phi^{\rm ex}$ peaks are only separated by 33~mas or
about 7~AU. In e$_4$ the 
A2$_{\rm N}$ and a$_{\rm N}$ structures look very
similar to those in e$_3$, but they moved together in
north direction with a proper motion of about 16~${\rm mas\,yr}^{-1}$. 

These observations can be interpreted as large, outward moving 
gas and dust cloud with a changing illumination from
the ionizing source. This produces for epoch e$_1$ a bright ionized
H$\alpha$ region on the west side of the dust clump. In e$_2$
much of the bright, dense
H$\alpha$ region A$_{\rm N}$ from e$_1$ has probably recombined, and 
only the emission from the filament A1$_{\rm N}$
remained, possibly because faint, lower density gas takes
longer to recombine. It
also seems that the A1$_{\rm N}$ structure is located behind the dust
cloud because its emission is weak or absent at the position
of the dust peak a$_{\rm N}$ as expected for a partly hidden
H$\alpha$ emission behind an opaque dust cloud.
The new, strong H$\alpha$ emission
A2$_{\rm N}$ appearing for epoch e$_3$ on the southeast side of the dust region
indicates that the emission direction of the ionizing radiation
from the hot component has changed.
The location and compactness of the A2$_{\rm N}$ emission strongly suggests
that the hot star ionizes the surface of a dense, dusty gas cloud
with a diameter of about 20~AU or 100~mas.

This dusty H$\alpha$ cloud is interesting, because its life time of
more than 7~years
in such a harsh environment exposed to
ionizing radiation and ionized gas from the hot component is
surprising. Apparently, dust erosion processes take place, as one can infer
from the faster moving dust tail features shown in Fig.~\ref{DustNorth}.
Strongly illuminated dust clouds such as a$_{\rm N}$
in R~Aqr could be interesting laboratories for investigations of
the complex chemical processes taking place in strongly illuminated
dusty outflows of cool stars \citep{VandeSande22,VandeSande24}.

\begin{figure}
\includegraphics[trim=0.cm 1.3cm 1.5cm 1.5cm,clip,width=8.7cm]{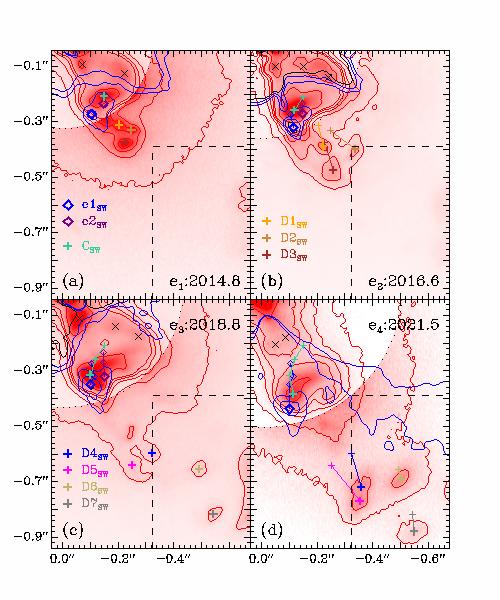}    
\caption{H$\alpha$ emission $I_{{\rm H}\alpha}$ in the inner region
  of the southwest outflow for the four VLT epochs.
  Relative $I_{{\rm H}\alpha}$ contours for flux levels
  $c\cdot n$ with $n=1,2,3,4,5,10$ are given in red. The color scale has a
  radial step of a factor of three marked by the dotted line to show
  bright clouds near the binary and fainter clouds further out.
  Blue contours and points illustrate
  the dust polarization in this region. Clouds are identified and replotted in later epochs
  with small symbols and lines to indicate cloud motions. Black
  crosses ($\times$) mark clouds discussed in Sect.~\ref{Hato03}.
  The dashed lines indicates the region also covered by Fig.~\ref{HaSW3}.}
\label{HaSW2}
\end{figure}

\begin{figure}
\includegraphics[trim=0.0cm 1.2cm 1.7cm 1.5cm,clip,width=8.8cm]{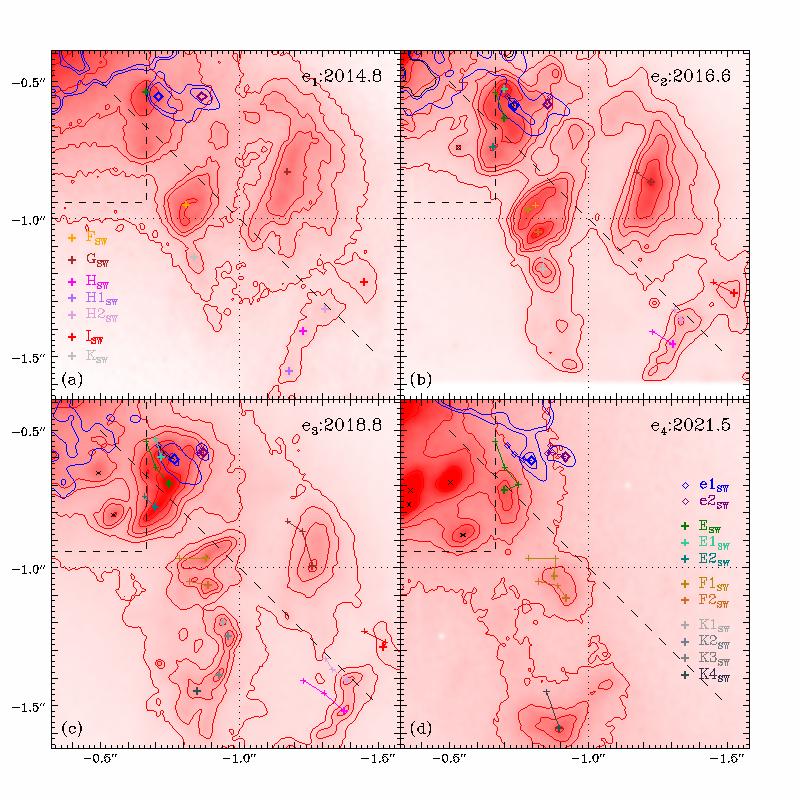}  
\caption{Same as Fig.~\ref{HaSW2} but for the H$\alpha$ clouds
  at larger separation. Relative $I_{{\rm H}\alpha}$ contours for flux levels
  $c\cdot n$ with $n=4,6,8,10,15,20$ are given in red. Blue contours
  and points illustrate
  the dust polarization. Clouds are identified and replotted in later epochs
  with small symbols and lines to indicate cloud motions.
  The dashed lines in the upper left
  show the region also covered by
  Fig.~\ref{HaSW2}. The diagonal dashed line marks the SW direction $\theta=225^\circ$. }
\label{HaSW3}
\end{figure}

\subsubsection{The H$\alpha$ emissions in the SW}

We take also a detailed look at the structure and evolution of the
H$\alpha$ and dust clouds in the SW outflow of R~Aqr. The innermost
region near the binary up to $r\approx 0.3''$ is shown in
Fig.~\ref{HaDustCent} and the clouds A$_{\rm SW}$ and B$_{\rm SW}$
were already described in Sect.~\ref{Hato03}. 
The regions from $r\approx 0.1''$ to $1.0''$
and $r\approx 0.5''$ to $2''$ are shown in Figs.~\ref{HaSW2} and
\ref{HaSW3}, respectively. Note
that the brightnesses of the H$\alpha$ clouds decrease strongly with
separation as quantified for epoch 2014.8 in Paper~I. Therefore, the maps in
Fig.~\ref{HaSW2} include a color scale step of a factor of three for
an $r$ expanding with epoch to display the structures of the much
brighter inner clouds together with the clouds at larger separation.

\paragraph{The H$\alpha$ cloud C$_{\rm SW}$.}
The cloud C$_{\rm SW}$ visible in Fig.~\ref{HaSW2}
moves during the four epochs from about $r=0.25''$ to $0.40''$
with a proper motion of about $22.3~{\rm mas\,yr}^{-1}$.
C$_{\rm SW}$ is a bright
cloud comparable in brightness to the feature
A$_{\rm N}$--A2$_{\rm N}$ in the north,
and it is also associated with dust scattering features c1$_{\rm SW}$ and
c2$_{\rm SW}$. The H$\alpha$ emission is closer to the central binary than the
polarization signal and therefore
also represents the illuminated side of a gas and dust cloud like
A$_{\rm N}$--A2$_{\rm N}$. Cloud C$_{\rm SW}$ has for epoch e$_1$ an elongated
structure with a tangential orientation, and in later epochs the
eastern part becomes dominant. This seems to be linked to the
overall illumination change, and therefore the apparent motion
of the cloud peak shows a tangential component toward the east.
The dust cloud is initially composed of the two components
c1$_{\rm SW}$ and c2$_{\rm SW}$. They move radially outward, and their
brightness distribution also shifts eastward like for H$\alpha$.
For the last epoch e$_4$ the component c2$_{\rm SW}$ disappeared or
merged with component c1$_{\rm SW}$. The polarization signal of c1$_{\rm SW}$
and c2$_{\rm SW}$ is comparable to the brightest cloud
on the east a$_{\rm E}$, but about four times fainter than a$_{\rm N}$
in the north. These three clouds have similar projected
separations to the Mira ($0.3''$ to $0.6''$), and one can expect roughly
a comparable illumination. This may indicate that there is less
dust in c1$_{\rm SW}$--c2$_{\rm SW}$ and therefore less
scattering polarization when compared to the bright dust cloud a$_{\rm N}$.
Interestingly, the fainter c1$_{\rm SW}$--c2$_{\rm SW}$
cloud also survived during the observed 6.7~years in the
middle of the strongly ionized SW jet outflow of R~Aqr.

\paragraph{Small clouds Dx$_{\rm SW}$.}
There are several compact and fainter clouds
D1$_{\rm SW}$ to D7$_{\rm SW}$ further out in Fig.~\ref{HaSW2}, outside of
the color map step. The recognition of a given cloud for the
different observing epochs is not always clear. The three clouds
D1$_{\rm SW}$, D2$_{\rm SW}$, and D3$_{\rm SW}$ show for the epochs
e$_1$ and e$_2$ a similar constellation, which just expanded by
about 80 to 100~mas to the southwest or with a proper motion of roughly
40 to 65~${\rm mas\,yr}^{-1}$.

After this it is unclear, whether one or both clouds of
D4$_{\rm SW}$ or D5$_{\rm SW}$ in the map e$_3$ are related to
the D3$_{\rm SW}$ in e$_2$, and whether the features
D4$_{\rm SW}$,D5$_{\rm SW}$ identified in e$_4$ are the same clouds
as in the previous epoch. 
We assume that D3$_{\rm SW}$ is not related to D4$_{\rm SW}$ and D5$_{\rm SW}$
and that D4$_{\rm SW}$-D5$_{\rm SW}$ in e$_3$ and e$_4$ are the same clouds.
In this case all five features D1$_{\rm SW}$ to D5$_{\rm SW}$ move with
a proper motion of roughly 45 to 60~${\rm mas\,yr}^{-1}$
and this is substantially faster
than the bright clouds A2$_{\rm N}$, B$_{\rm SW}$, and C$_{\rm SW}$.
Further support for faster moving ionized gas comes from the simultaneous
expansion of the outermost H$\alpha$ contour
in the four panels of Fig.~\ref{HaSW2},
which looks similar to an expanding bubble. 

The higher velocity of the five clouds D1$_{\rm SW}$ to D5$_{\rm SW}$
could be caused by a density drop in the outflow coinciding with the
color scale step in Fig.~\ref{HaSW2}. The brightness
of H$\alpha$ clouds is a good indicator
of the cloud density, because the H$\alpha$ emissivity scales
with the density squared (Paper~I). A density drop
can be produced by gas collisions 
and this could have triggered an expansion and split up of a larger
cloud before epoch e$_1$ into the three clouds D1$_{\rm SW}$,
D2$_{\rm SW}$, and D3$_{\rm SW}$, which are then accelerated
by a pressure expansion into the rarefied pre-shock gas.
The higher velocity for these clouds with
respect to the ambient medium could then introduces drag forces and
explain the fast dispersion and short life time 
for small clouds. The transient component A$'_{\rm N}$ in the north
seems to be a similar case (Fig.~\ref{HaDustCent}).
This scenario is inspired by the well established
clumpy stellar wind models for hot, luminous stars
\citep[e.g.,][]{Sundqvist18,Runacres02,Feldmeier95}
but it is unclear, whether such models are also applicable to the slow, dense
outflow in R~Aqr. In any case, R~Aqr is a unique object to study
the hydrodynamics of ionized, clumpy outflows because the evolution
of individual H$\alpha$ clouds can be followed with high
resolution imaging.

The small clouds D6$_{\rm SW}$ and D7$_{\rm SW}$ appear 
in epoch e$_3$ and they are again seen in the e$_4$ map.
They move slowly in a manner similar to the bright H$\alpha$ clouds B$_{\rm SW}$ and C$_{\rm SW}$
and might be just smaller, irradiated, and therefore ionized slow
moving clouds of the Mira wind. It is unclear if they also have a longer life time or whether they are transient, but clearly
not all compact H$\alpha$ clouds move fast.

\paragraph{H$\alpha$ clouds outside of 150~AU.}
The evolution of the clouds at separations of about $0.7''$ to
$2.0''$ or about 150~AU to 450~AU is shown in Fig.~\ref{HaSW3}.
Many of the clouds at this separation are more extended and have
a much lower surface brightness than
the clouds at smaller separation (Paper~I) as expected for
a freely expanding wind. The clouds show also quite complex
substructures and often significant structural changes with time.
For example some parts of cloud E2$_{\rm SW}$ become clearly
brighter and other parts fainter with time, and similar for G$_{\rm SW}$.
The cloud F$_{\rm SW}$ splits up into two components and we
derive proper motions only for F1$_{\rm SW}$ and F2$_{\rm SW}$ after the clear
split-up into two components. The two subcomponents
K3$_{\rm SW}$ and K4$_{\rm SW}$ seem to merge from epoch e$_3$ to e$_4$
into a single K4$_{\rm SW}$, and this is also supported by the HST map from
2020.0 in Fig.~\ref{FiveFigHST}. We use the K4$_{\rm SW}$ positions to
derive the proper motion.

The dominant change for the SW outflow
is the disappearance of the clouds E1$_{\rm SW}$, G$_{\rm SW}$,
H2$_{\rm SW}$, H3$_{\rm SW}$, and I$_{\rm SW}$ located on the west side or near
the $\theta=225^\circ$ line for the 2021.5 epoch as already
emphasized in Sect.~\ref{SectHaObs}. This general trend is already
visible between epoch 2016.6 and 2018.8.
for the prominent cloud complex G$_{\rm SW}$ located far in the west at
$\theta\approx 230^\circ - 240^\circ$.

An interesting cloud is the H$\alpha$
emission complex E$_{\rm SW}$ which includes the dust scattering
clouds e1$_{\rm SW}$ and perhaps also e2$_{\rm SW}$.
The H$\alpha$ emission E$_{\rm SW}$ is also located on the illuminated
side of the dust cloud e1$_{\rm SW}$ separated in 2014.8 by only 45~mas or 10~AU.
The H$\alpha$ emission E$_{\rm SW}$ changes strongly with time, first
from a circular structure with a southern extension in
epoch e$_1$, to a much brighter, elongated and structured cloud
for epochs e$_2$ and e$_3$, and then a much fainter feature on the
south-east of the dust for e$_4$ during the eclipse phase.
For that epoch the H$\alpha$ emission peak is separated
by more than $145$~mas or 32~AU from e1$_{\rm SW}$.
Cloud e2$_{\rm SW}$ is about $120$~mas further west of
cloud e1$_{\rm SW}$, and it is not clear whether the two 
features are associated with each other. However, the two dust
clouds remained during all these H$\alpha$ variations for E$_{\rm SW}$
practically unchanged and just continued their slow radial motion.
This indicates that the presence or absence of ionizing radiation
from the hot component has no obvious
impact on the polarized scattered light from the dust cloud e1$_{\rm SW}$
illuminated by the Mira.

\begin{figure}
\includegraphics[trim=2.5cm 13.cm 9.1cm 2.9cm,width=8.8cm]{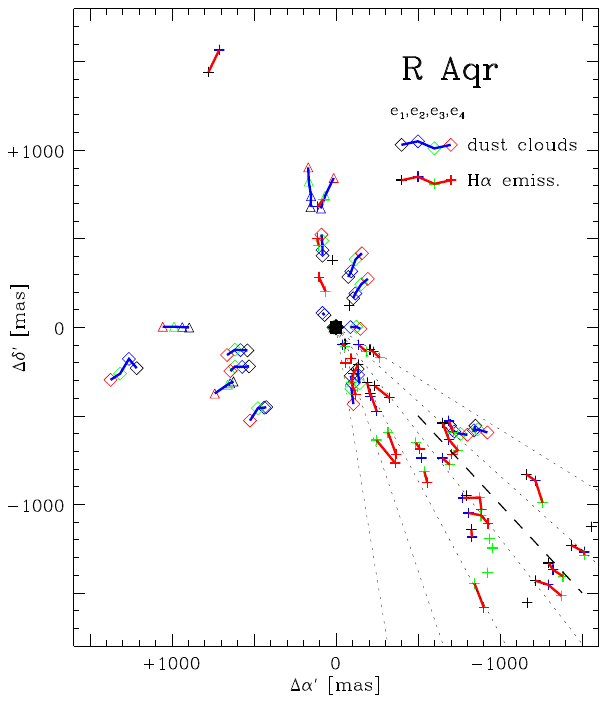}
\caption{Position map for the dust clouds (diamonds), dust fronts and
  tail features (triangles), and H$\alpha$ emission
  clouds (crosses) for the epochs e$_1$ (black), e$_2$ (blue), e$_3$ (green),
  and e$_4$ (red). Blue and red lines indicate
  the motion of the apparently same clouds between different epochs. The
  dotted lines define angular bins of $\Delta\theta=10^\circ$ for the
  SW outflow, and the SW direction ($\theta=225^\circ$) is given by
  the dashed line.}
\label{PMMap}
\end{figure}

\section{Distribution and motion of dust and H$\alpha$ clouds}
\label{SectPMall}

The figures in the previous sections show the local peak positions
$Q_\phi^{\rm ex}$ for the dust clouds and
$I_{{\rm H}\alpha}$ for the H$\alpha$ emissions
for the four epochs e$_1$, e$_2$, e$_3$, and e$_4$.
Most clouds are seen for different epochs, which
allows a determination of their proper motions. There are some
uncertainties because of cloud shape changes, in particular for
the H$\alpha$ emissions, because of an evolution in their density
structure or illumination shifts, so that the flux maxima may not
represent accurately the ``physical'' mass centroid of a cloud.
Despite this, we base the motion measurement strictly on
local flux maxima serving as clear definition for the cloud positions.

All cloud positions ($\Delta \alpha,\Delta \delta$) in the maps
are given relative to the Mira, because this is a very
reliable reference point. However, the center of mass is a better
reference not affected by orbital motion, and its position can be
derived from the positions of the hot component
($\Delta \alpha_{\rm hot},\Delta \delta_{\rm hot}$)
given in Table~\ref{TabObs} \citep[from][]{Alcolea23} and
assuming a mass ratio $m_{\rm Mira}/m_{\rm hot}=2$. The
cloud positions relative to the center of mass are then 
$\Delta \alpha' = \Delta \alpha + \Delta \alpha_{\rm hot}/3$ and
$\Delta \delta' = \Delta \alpha + \Delta \delta_{\rm hot}/3$. 
This correction changes for our data mainly the radial motion of the
dust clouds on the east side by about $-3~{\rm mas\,yr}^{-1}$ and on the
west side by $+3~{\rm mas\,yr}^{-1}$.

\begin{table}
  \caption{Position and radial motion of dust structures in R~Aqr.}  
\label{TabDustMotion}
\begin{tabular}{lcccccc}
\noalign{\hrule}
\noalign{\smallskip\hrule\smallskip}
name         & $r'_f$ &  $\theta'_f$ & $e_f,e_l$ & ${\rm pm}'_r$ & $t_{\rm dyn}$
                                                             \cr
             & mas      &  degr    &          & mas/yr       &   yr   \cr
\noalign{\smallskip\hrule\smallskip}
\noalign{\noindent dust clouds \smallskip}
a$_{\rm N}$   & 412    & 11         & 1,4      & $17.5\pm1.5$ &  24    \cr
a$_{\rm NE}$  & 98     & 45         & 1,2      & $9.6\pm5.6$  &  10    \cr
a$_{\rm E}$   & 556    & 104        & 1,4      & $18.8\pm1.5$ &  30    \cr
b$_{\rm E}$   & 574    & 113        & 1,4      & $17.1\pm1.5$ &  36    \cr
c$_{\rm E}$   & 1236   & 101        & 1,4      & $25.0\pm1.5$ &  49    \cr
a$_{\rm SE}$  & 620    & 137        & 1,4      & $18.3\pm1.5$ &  34    \cr
c1$_{\rm SW}$ & 288    & 198        & 1,4      & $23.5\pm1.5$ &  12    \cr
c2$_{\rm SW}$ & 270    & 209        & 1,3      & $19.3\pm2.5$ &  14    \cr
e1$_{\rm SW}$ & 890    & 231        & 1,4      & $17.0\pm1.5$ &  52    \cr
e2$_{\rm SW}$ & 1016   & 236        & 1,4      & $12.6\pm1.5$ &  84    \cr
a$_{\rm W}$   & 88     & 271        & 2,4      & $12.3\pm2.0$ &  7     \cr
a$_{\rm NW}$  & 194    & 372        & 1,4      & $20.8\pm1.5$ &  9     \cr
b$_{\rm NW}$  & 294    & 345        & 1,4      & $22.7\pm1.5$ &  13    \cr
\noalign{\smallskip}
&        & \multispan{2}{\hfil $\langle {\rm pm}'_r \rangle \pm \sigma_{\rm pm}$ :}
                                             & $18.0\pm 4.5$ &     \cr

\noalign{\smallskip \noindent dust fronts \smallskip}
s$_{\rm E}$  & 895     & 90       & 1,4     & $23.9\pm3.0$  &  38    \cr
s$_{\rm SE}$ & 695     & 116      & 1,4     & $19.6\pm3.0$  &  35    \cr

\noalign{\smallskip \noindent dust tail features \smallskip}
t1$_{\rm N}$  & 696   & 13       & 1,4     & $32.9\pm3.0$  &  21    \cr
t2$_{\rm N}$  & 677   &  8       & 2,4     & $33.3\pm4.0$  &  20     \cr
\noalign{\smallskip\hrule\smallskip}
\end{tabular}
\tablefoot{Radial proper motion (pm$'_r$) relative to the center of mass
  and position in polar coordinates ($r'_f,\,\theta'_f$) for the first
  epoch (e$_f$) that a given feature was detected. We note that e$_l$ is 
  the last epoch considered for the derivation of ${\rm pm}'_r$ 
  and the dynamical timescale $t_{\rm dyn}$.
  }
\end{table}

The corrected cloud positions and motions are illustrated in
the chart in Fig.~\ref{PMMap}
and the derived radial components for the proper motions 
relative to the center of mass ${\rm pm}'_r$ for the dust clouds and
the H$\alpha$ emissions are given in the Tables~\ref{TabDustMotion}
and \ref{TabHaMotion}, respectively.
The ${\rm pm}'_r$ values follow from the positions
$\Delta \alpha',\,\Delta \delta'$ for the first $e_f$ and last $e_l$
epoch when the flux peak of a given cloud is observed. The indicated errors
for ${\rm pm}'_r$ assume an uncertainty of $\Delta r=\pm 7$~mas
for the position of a cloud peak
and $\pm 10$~mas for the radial position change between epochs
$e_l$ and $e_f$. This yields an uncertainty
$\Delta {\rm pm}_r=10~{\rm mas}/\Delta(t)=1.5$~mas/yr
for a cloud observed for all four epochs or $\Delta t=e_l-e_f=6.7$~yr.
These errors do not include cloud motion uncertainties introduced by
illumination or cloud structure changes.

The tables list also the cloud dynamical timescale
$t_{\rm dyn} = r(e_f)/\Delta {\rm pm}'_r$ 
assuming a formation very close to the center and a
constant velocity. Projected sky velocities can be calculated
with the relation
$v_{\rm sky}\,[{{\rm km\,s}}^{-1}] =
4.74\cdot {\rm pm}'_r\,[{{\rm arcsec\,yr}}^{-1}]\cdot d\,{\rm [pc]}$,
or for the adopted distance of $d=218$~pc of R~Aqr with the conversion factor
\begin{equation}
v_{\rm sky}\,[{\rm km\,s}^{-1}] = 1.03 \cdot {\rm pm}'_r\,[{\rm mas\,yr}^{-1}]\,.
\label{FactConv}
\end{equation}

\subsection{Dust cloud velocities}
\label{SectPMdust}
\begin{figure}
\includegraphics[trim=2.0cm 15cm 3.0cm 3.2cm, width=8.5cm]{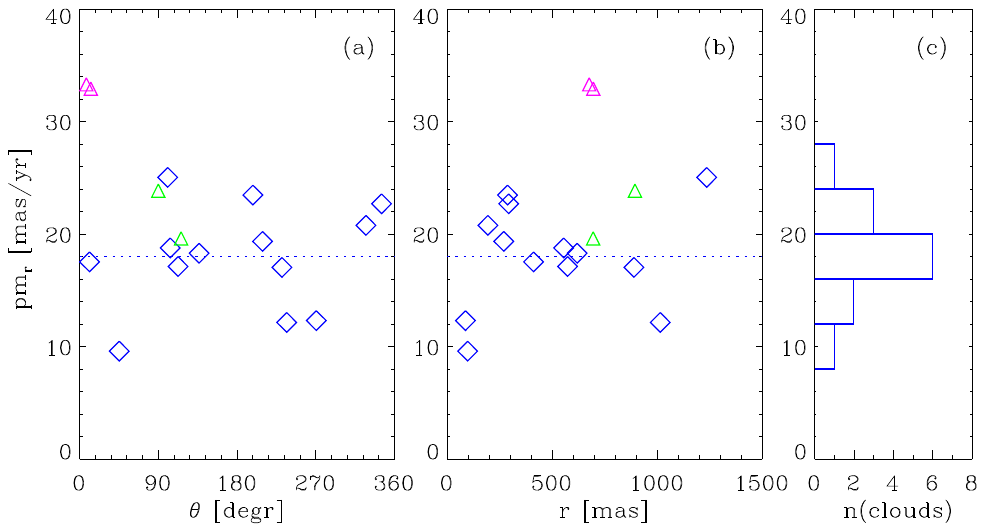}
\caption{Proper motion versus position angle $\theta$ (panel a)
  and separation $r$ (b) for the dust clouds (blue),
  dust polarization steps (magenta), and the dust tail features (green).
  The histogram (c) shows the velocity distribution
of the clouds in $4~{\rm mas\,yr}^{-1}$ wide bins.}
\label{DiagPMDust}
\end{figure}

The dust clouds are typically observed for all four epochs,
and they show a slow, steady radial motion.
The diagrams in Fig.~\ref{DiagPMDust} illustrate the spatial
distribution and proper motions pm$'_r$ of the dust clouds.
There are no obvious angular or radial dependencies for the 
pm$'_r$-values
and the mean value for the clouds is
$\langle {\rm pm}'_r\rangle=18.0~{\rm mas\,yr}^{-1}$ with standard
deviation $\sigma_{\rm pm}=4.5~{\rm mas\,yr}^{-1}$ (Table~\ref{TabDustMotion}).
An interesting aspect are the slow motions of the
two innermost dust clouds a$_{\rm NE}$ and a$_{\rm W}$. It is expected that
stellar winds start slowly near the star, but the available evidence
for such an wind acceleration region is not strong and should
be based on a larger sample of clouds at small separation $r<0.1''$
or $<20$~AU.

Moving dust density steps and tail features are also included
in Fig.~\ref{DiagPMDust} but they are not considered in the histogram
in panel (c). Larger uncertainties for pm$'_r$ are assumed
for these features (Table~\ref{TabDustMotion}) because they are less
well defined than the clouds.
The step features s$_{\rm E}$ and s$_{\rm SE}$ move roughly with the speed of the
dust clouds. The two dust tail features t1$_{\rm N}$ and t2$_{\rm N}$ of the
northern cloud have a significantly larger radial velocity
(Sect.~\ref{SectCometDust}).
This could be caused by radiation pressure from the Mira
and an enhanced acceleration for certain kinds of
dust particles and perhaps also by drag forces of the
ionized and possibly magnetized outflow from the hot component exerted on
charged dust (Sect.~\ref{SectCometDust}).

The estimated dynamical time $t_{\rm dyn}$ of the clouds range from
7 to 84 year with a typical value of about 30~years. Practically
all clouds at projected separations larger than
150~mas (30~AU) could be recognized for all four epochs or for a
time span of 6.7~years indicating that dust cloud or dust
concentration, once formed, persist over decades in the outflow of R~Aqr.
This is even true for the clouds a$_{\rm N}$,
c1$_{\rm SW}$, and e1$_{\rm SW}$, which are exposed to the energetic
radiation and ionized outflow from the hot component.

Interestingly, the dust clouds and the dust steps on the
east and southeast have on average $t_{\rm dyn}\approx 37\pm 6$~yr
with respect to the reference epoch ${\rm e}_1=e_f$. This suggest that they
were formed around $1977 \approx {\rm e}_1-\langle t_{\rm dyn}\rangle$
or almost one orbit ago. The presence of the prominent dust
features on the east side could be related to a spiral
pattern in the orbital plane of R~Aqr as described in \cite{KimH24}
and be related to the orbital motion of the Mira and the last periastron
passage of the dust enshrouded hot component
(see also Sect.~\ref{SectDustEast}). 

The mean motion $\langle {\rm pm}'_r\rangle$ of the dust clouds
yields according to Eq.~\ref{FactConv} 
$\langle v'_{\rm d,sky}\rangle = 18.6~{\rm km\,s}^{-1}$
for the dust velocity component parallel to the sky plane.
The full outflow velocity $v'_d$ is composed of $v'_{\rm d,sky}$ and
a radial component $v'_{\rm d,RV}$, and there is
$v'_{\rm d,sky}=v'_d \sin \phi$, where $\phi$ is the angle between the
cloud propagation direction and the line of sight. Polarimetry
favors the detection of clouds producing a strong polarization
and this is expected for scattering angles Mira -- cloud -- observer
of roughly $\phi\approx 45^\circ$ to $135^\circ$.
Scattering clouds in front or behind the Mira can produce a strong intensity
signal but the polarization for forward $\phi\approx 0^\circ$
and backward $\phi\approx 180^\circ$ scattering is low, and
therefore such clouds
may not be detected in the $Q_\phi^{\rm ex}$ maps.
The scattering polarization depends on
the dust particles and the optical depths of the cloud, and according to
\cite{Zubko00} the detected clouds with substantial polarization
have scattering angles of roughly $\phi \approx 90^\circ\pm 45^\circ$. Thus,
the observed velocity components $v'_d \sin \phi$ represent
between 70~\% and 100~\% of the full outflow velocity $v'_{\rm d}$.
Based on this we can estimate a mean outflow velocity for the
sampled dust clouds according to
$\langle v'_d \rangle \approx \langle v'_{\rm d,sky}\rangle/0.85
= 21.9~(\pm 2.2)~{\rm km\,s}^{-1}$.

\paragraph{Comparison with molecular outflow velocities.}
The mean velocity for the dust outflow $\langle v'_d \rangle$
can be compared to the RV width of molecular lines of R~Aqr. ALMA observation
show CO emissions with
$\Delta{\rm RV_{CO}}\approx \pm 17~{\rm km\,s}^{-1}$
with respect to the systemic RV of $v_{\rm sys}\approx -24~{\rm km\,s}^{-1}$
and practically no detection for larger $|\Delta{\rm RV}|$ 
\citep{Ramstedt18,Bujarrabal18,Bujarrabal21,GomezG24}. 
The CO velocity maps indicate some rotational motion of the gas
in the orbital plane with relative blue shifts on the west and red shifts
on the east of the binary.

The estimated dust outflow velocity 
$\langle v'_d \rangle\approx 21.9~{\rm km\,s}^{-1}$ is larger than
$|\Delta{\rm RV_{\rm CO}}|\approx 17~{\rm km\,s}^{-1}$ for the
CO gas, and this is expected for dust-driven winds accelerated
by radiation pressure, where drag forces between dust and gas
also accelerate the gas, but to a lower velocity 
than for the dust \citep[e.g.,][]{Goldreich76,Lamers99,Mastrodemos96}.
The derived observational ratio 
$\Lambda=\langle v'_d \rangle/|\Delta{\rm RV_{CO}}|\approx 1.3$
for the relative drift velocity between dust and gas
in R~Aqr is a very useful parameter for the modeling of dust-driven winds. 

Dust proper motions and gas RVs measure different velocity
components and outflow regions. For the interpretation
of the $\Lambda$ ratio for R~Aqr system one needs to consider
the strong deviations from spherical symmetry of the dust and CO maps, and
the CO velocity profile \citep{Ramstedt18}.
In addition, the conversion
$\langle {\rm pm}'_r \rangle$ $\rightarrow$ $\langle v'_{\rm d,sky}\rangle$
depends on the adopted distance to R~Aqr.
The used value of $d=218~$pc is at the lower end of the existing
distance estimates \citep[e.g.,][]{Alcolea23,Liimets25}, and
larger $d$ would give higher
dust cloud velocities. However, future GAIA data should provide a
more accurate distance and additional mapping of the dust and gas for
R~Aqr will also better constrain the 3D outflow geometry and provide
better knowledge on the dust and gas velocities.
Perhaps, it is also possible in the future to measure
simultaneously proper motions of the dust scattering clouds
and of the CO gas emission for a direct measurement of the relative
drift velocity between dust and CO gas.

Several bright AGB stars also show evolving
circumstellar polarization structures in SPHERE/ZIMPOL
observations like W Hya, R Dor, and others
\citep[e.g.,][]{Khouri16,Khouri20,Khouri26,Ohnaka17,Montarges23,Badolo26}.
However, up to now no steadily expanding patterns of well
defined, compact dust
clouds, like in R~Aqr, were reported. Observing the expansion of
such a cloud pattern in combination with CO data for more symmetric
outflows around single AGB stars would be ideal to derive
the dust to gas velocity ratio $\Lambda$. One reason for the lack of such
cloud detections could be the lower sensitivity for
most AGB star data, because the 
observations focussed more on the bright inner scattering region
$r\lapprox 200$~mas. Close to the star the polarization shows
a steep gradient and faint dust clouds may not produce
a local polarization maximum. It is also possible that the formation
of many compact and prominent clouds in a dusty wind is a
special feature of the strongly interacting R~Aqr system.

\begin{figure}
  \includegraphics[trim=2.0cm 15.0cm 3.5cm 3.0cm,clip,width=8.8cm]{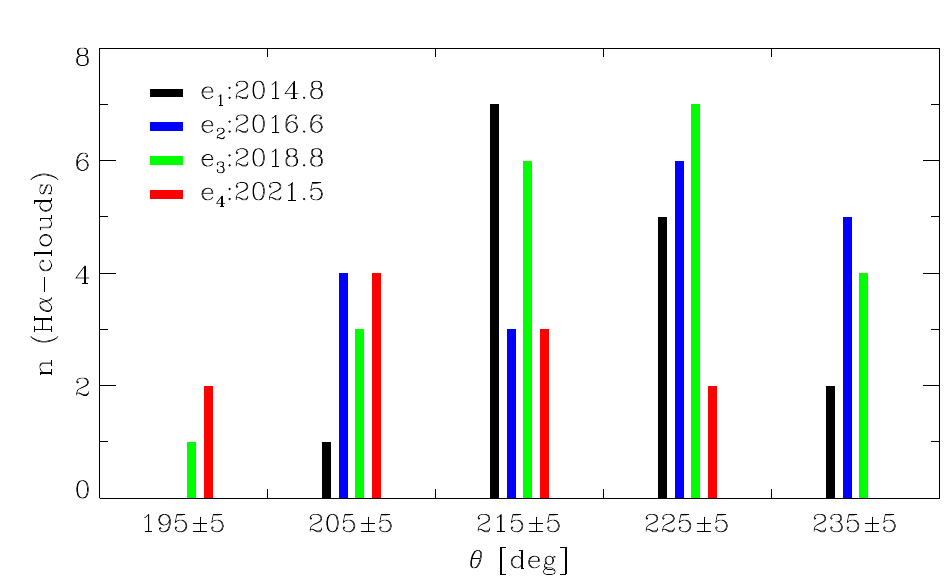}  
  \caption{Distribution of H$\alpha$ clouds for the four epochs in
    angular bins of $\Delta\theta=10^\circ$ in the SW outflow region.}
\label{DiagHaSWpos}
\end{figure}

\subsection{Distribution of the H$\alpha$ clouds and illumination shift}

Most H$\alpha$ clouds are located in SW wedge between 
$\theta = 195^\circ$ and $239^\circ$ and some in the N-NE
between $3^\circ$ and $28^\circ$ representing
the two opposite, collimated outflows (Fig.~\ref{PMMap}).
An important phenomenon for the SW outflow
is the obvious disappearance of the H$\alpha$
clouds between epoch e$_3$ and e$_4$ for position angles
$\theta>225^\circ$ (e.g., Figs.~\ref{FourFigHaVLT},\ref{FiveFigHST}).
To quantify this, we count the H$\alpha$ cloud for all four epochs in the
$\Delta\theta = 10^\circ$ bins illustrated in Fig.~\ref{PMMap}.
The resulting distributions given in Fig.~\ref{DiagHaSWpos} are
very similar for the first three epochs with a mean value
of about $\langle \theta \rangle=220^\circ$. For the
last epoch 2021.5 there are no H$\alpha$ clouds in the
$\theta=235^\circ \pm 5^\circ$ bin and the mean value for the
cloud distribution has shifted to about $\langle \theta \rangle=210^\circ$.
The concurrent disappearance of the clouds with $\theta>225^\circ$
at all separations from $r\approx 30~$AU to 300~AU strongly
suggests the occurrence of an angular shift of the illumination
by ionizing radiation from the hot component. This happens
together with the dust eclipse event and is
probably caused by a rearrangement of dust near the
ionizing source. The disappearance of the H$\alpha$ 
emission within a few years indicates a typical recombination time
scale of $\lapprox 1$~yr as expected for clouds with
electron densities larger than $N_e\gapprox 10^5 {\rm cm}^{-3}$
(Paper I).  

Shocks from an energetic jet outflow were often proposed as main ionization
mechanism for the collimated outflow in R~Aqr. However,
the simultaneous disappearance of clouds
at small and large separation seems to exclude this idea, because
a shift or switch-off of a jet stream would produce a delay
for the cloud recombination around $r\approx 300$~AU with respect
to $r\approx 30$~AU of more
than a few years for a jet velocity of $v_{\rm jet}<500~{\rm km\,s}^{-1}$
($<500~{\rm mas\,yr}^{-1}$). There exist no proper motion measurements
of R~Aqr pointing to gas velocities larger than
$250~{\rm km\,s}^{-1}$, and for separations
covered by the ZIMPOL data, this limit is around $\approx 150~{\rm km\,s}^{-1}$
\citep{Melnikov18,Liimets18}. Nonetheless, shocks
can still be present in the collimated outflow and contribute
to the ionization and gas heating in particular for the
low density gas. This component fills about 90~\% of the volume 
of the ionized outflow region (Paper I), and which
produces probably the faint diffuse emission seen in the sensitive
HST H$\alpha$ data presented in \cite{Melnikov18}.

\begin{table}
  \caption{Position and motion of the H$\alpha$ clouds.}  
\label{TabHaMotion}
\begin{tabular}{lcccccc}
\noalign{\hrule}
\noalign{\smallskip\hrule\smallskip}
name         & $r'_f$ &  $\theta'_f$ & $e_f,e_l$ & ${\rm pm}'_r$ & $t_{\rm dyn}$
                                                             & note \cr
             & mas      &  degr    &          & mas/yr       &   yr   \cr
\noalign{\smallskip\hrule\smallskip}

\noalign{\smallskip \noindent H$\alpha$ clouds \smallskip}
\noalign{\smallskip \noindent N-NE outflow \smallskip}
A$_{\rm N}$   &  380    &  3         & 1,1      &              &     & e,d \cr
A1$_{\rm N}$  &  690    &  9         & 2,3      & $16.4\pm4.4$ &  42 & e,d \cr
A2$_{\rm N}$  &  475    & 13         & 3,4      & $14.9\pm3.7$ &  32 & e,d \cr
A$'_{\rm N}$  &  213    & 17         & 3,4      & $32.5\pm3.7$ &  7  & c  \cr
C$_{\rm NE}$  & 1637    & 28         & 1,2      & $47.0\pm5.6$ &  35 & c \cr
``Hinge''    &  148    & 327        & 1,1      &              &     & c,d\cr
\noalign{\smallskip \noindent SW outflow \smallskip}
A$_{\rm SW}$  &  109    & 210        & 1,3      & $2.7\pm2.5$\tablefootmark{a}
                                                             &    & e \cr
A1$_{\rm SW}$ &  210    & 195        & 4,4      &              &    & e \cr
A2$_{\rm SW}$ &  199    & 207        & 4,4      &              &    & e \cr
B$_{\rm SW}$  &  242    & 239        & 1,3      & $18.7\pm2.5$ &  13 & e  \cr
B1$_{\rm SW}$ &  170    & 234        & 2,3      & $26.5\pm4.4$ &  6  & e  \cr
C$_{\rm SW}$  &  250    & 212        & 1,4      & $22.3\pm1.5$ &  11 & e,d \cr
D1$_{\rm SW}$ &  366    & 211        & 1,2      & $42.9\pm5.6$ &  9  & c    \cr
D2$_{\rm SW}$ &  406    & 215        & 1,2      & $61.8\pm5.6$ &  7  & c    \cr
D3$_{\rm SW}$ &  431    & 209        & 1,2      & $58.9\pm5.6$ &  7  & c    \cr
D4$_{\rm SW}$ &  673    & 208        & 3,4      & $48.3\pm3.7$ &  14 & c   \cr
D5$_{\rm SW}$ &  683    & 201        & 3,4      & $60.5\pm3.7$ &  11 & c   \cr
D6$_{\rm SW}$ &  814    & 217        & 3,4      & $15.5\pm3.7$ &  52 & c   \cr
D7$_{\rm SW}$ &  904    & 215        & 2,4      & $27.1\pm2.0$ &  33 & c    \cr
E$_{\rm SW}$  &  846    & 230        & 1,3      & $23.8\pm2.0$ &  36 & e,d  \cr
E1$_{\rm SW}$ &  868    & 232        & 2,3      & $27.4\pm4.4$ &  32 & e,d  \cr
E2$_{\rm SW}$ &  983    & 221        & 2,3      & $26.2\pm4.4$ &  38 & e,d  \cr
\noalign{\smallskip}
F$_{\rm SW}$ &   1240   & 220        & 1,1      &               &      & e   \cr
F1$_{\rm SW}$ &  1236   & 219        & 2,4      & $ 24.2\pm2.0$ & 51 & e   \cr
F2$_{\rm SW}$ &  1324   & 218        & 2,4      & $ 23.9\pm2.0$ & 55 & e\cr
G$_{\rm SW}$  &  1427   & 234        & 1,3      & $43.8\pm2.5$ &  33   & e\cr
H$_{\rm SW}$  &  1877   & 220        & 1,3      & $44.2\pm2.5$  &  45  & e \cr
H1$_{\rm SW}$ &  1944   & 217        & 1,1      &              &       & e\cr    
H2$_{\rm SW}$ &  1857   & 224        & 1,3      & $29.0\pm2.5$  &  64  & e \cr
I$_{\rm SW}$  &  1891   & 229        & 1,3      & $24.5\pm2.5$ &  79   & e\cr
J$_{\rm SW}$  &  1923   & 234        & 1,1      &               &      & e\cr
K$_{\rm SW}$  &  1408   & 216        & 1,2      & $20.8\pm5.6$  & 68   & c \cr
K1$_{\rm SW}$ &  1517   & 218        & 3,3      &               &      & e \cr
K2$_{\rm SW}$ &  1570   & 217        & 3,3      &               &      & e \cr
K3$_{\rm SW}$ &  1667   & 214        & 3,3      &               &      & e \cr
K4$_{\rm SW}$ &  1675   & 210        & 2,3      & $53.3\pm3.7$  & 31   & e \cr
\noalign{\smallskip}
&  & \multispan{2}{\hfil $\langle {\rm pm}'_r \rangle \pm \sigma_{\rm pm}$ :}
                                             & $33.6\pm 16.3$ &     \cr

\noalign{\smallskip\hrule\smallskip}
\end{tabular}
\tablefoot{The parameters $r'_f$, $\theta'_f$, $e_f$, $e_l$, ${\rm pm}'_r$,
  and $t_{\rm dyn}$ are identical to the definition given in
  Table~\ref{TabDustMotion}. The letters in the column ``note'' characterize
  the H$\alpha$ clouds: c = compact, e = extended, d = with polarization from
  dust. \tablefoottext{a}{No dynamical timescale derived for this
  quasi-stationary feature.}}
\end{table}

\subsection{Proper motions for the H$\alpha$ clouds}

The H$\alpha$ clouds observed with ZIMPOL show a large range
in radial proper motion from
${\rm pm}'_r= 15~{\rm mas\,yr}^{-1}$ to $62~{\rm mas\,yr}^{-1}$ with
a mean of $\langle {\rm pm}'_r\rangle = 34~{\rm mas\,yr}^{-1}$ and
a standard deviation $\sigma_{\rm pm}= 16~{\rm mas\,yr}^{-1}$
(Table~\ref{TabHaMotion}). Beside this, they also show
a large diversity in size, structure and life time. Only
few H$\alpha$ emission features are observed for all epochs,
while many clouds are only seen in three,
two, or even only one epoch because they undergo significant
brightness changes.

\begin{figure}
  \includegraphics[trim=2.5cm 13.0cm 8.8cm 3.5cm,clip,width=8.5cm]{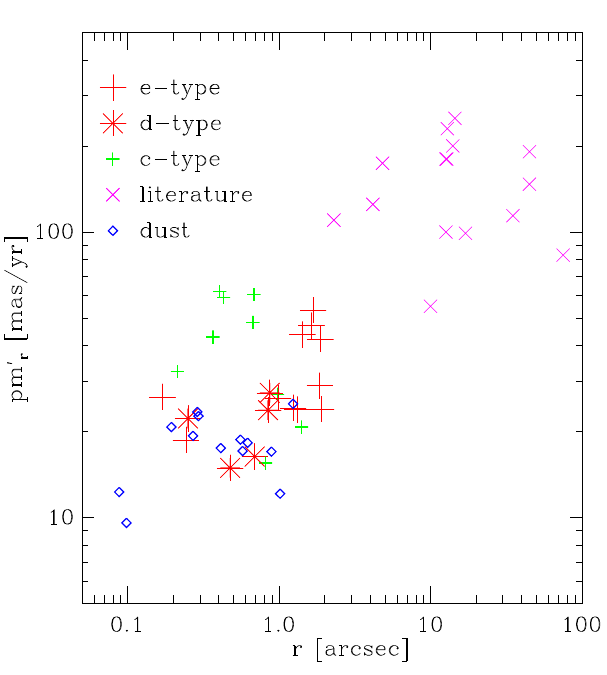}  
  \caption{Proper motion pm$'_r$ of the H$\alpha$ clouds as function
    of separation using different symbols for the three H$\alpha$ cloud
    classes at $r<2''$ and literature values are included for $r>2''$. The
    dust clouds from Fig.~\ref{DiagPMDust} are also plotted.}
\label{DiagPMHalpha}
\end{figure}

We distinguish for the analysis of the proper motion 
between groups of H$\alpha$ clouds with different properties
as indicated in column ``notes'' in Table~\ref{TabHaMotion}:
\begin{itemize}
\item{e:} extended H$\alpha$ clouds with diameters $d\gapprox 100$~mas,
  showing substructures like C$_{\rm SW}$ and G$_{\rm SW}$,
  or groups of associated clouds, like the clouds
  A$_{\rm N}$--A1$_{\rm N}$--A2$_{\rm N}$ or clouds B$_{\rm SW}$--B1$_{\rm SW}$,
  and others. These clouds have a lifetime of
  $>4$~yr and are present for all four epochs or for the
  first three epochs if they are located in the SW outflow to the west of the
  $\theta=225^\circ$ line and disappear for the last epoch e$_4$
  because of the illumination shift.
\item{d:} dusty H$\alpha$ clouds showing 
  scattering polarization, like A2$_{\rm N}$, C$_{\rm SW}$, or E$_{\rm SW}$,
  with their H$\alpha$ emission located
  on the illuminated side of the scattering dust clouds, a$_{\rm N}$,
  c1$_{\rm SW}$--c2$_{\rm SW}$, or e1$_{\rm SW}$, respectively.
  The d-type clouds are also extended (e-type) clouds
  except for the ``Hinge'' (see Fig.~\ref{HaDustCent}a), which resembles
  more of a transient arc-like extension to the central source.
  No d-type clouds are detected for separations $r\gapprox 1''$.
\item{c:} compact clouds have a small diameter $d<100$~mas
  and a short lifetime, because they are only seen during
  one or two epochs. Good examples are A$'_{\rm N}$ or the clouds
  D1$_{\rm SW}$, D2$_{\rm SW}$, and D3$_{\rm SW}$.
\end{itemize}
This is only a rough scheme with a few uncertain cases, but for most
clouds the classification is obvious. The measured proper motions
pm$'_r$
are plotted in Fig.~\ref{DiagPMHalpha} as function of
separation $r'_f$. Different colors are used for
c-type (green) H$\alpha$ clouds and e-type clouds (red), while
e-type clouds with dust emission (d-types) are plotted with
star symbol instead of only a cross.
Also included are measurements from the literature for separations of
$r>2''$ given by \cite{Huang23} for the bubble
(H$_{\rm SW}$+I$_{\rm SW}$+J$_{\rm SW}$) visible in the lower right
corners in Fig.~\ref{FiveFigHST} and by \cite{Melnikov18} and
\cite{Liimets18} for H$\alpha$ features at much larger
separations. For comparison, the scattering dust clouds
from Table~\ref{TabDustMotion} are also plotted.

We focus first on the red symbols in
Fig.~\ref{DiagPMHalpha} for the bright, extended clouds at $r\lapprox 2''$.
The d-type H$\alpha$ clouds are all at $r<1''$ and move roughly
with the same low proper motion of about 20~mas\,yr$^{-1}$ like the
dust scattering clouds. Also extended clouds at smaller separation
$r<1.2''$ without dust scattering, like B$_{\rm SW}$--B1$_{\rm SW}$ and
F1$_{\rm SW}$--F2$_{\rm SW}$, move with such a
low velocity. This indicates that the bright innermost H$\alpha$
emission regions are ionized gas clouds from the red giant
wind.
Interestingly, the extended clouds
at separations of $r\approx 1.2''$ to $2''$ show often a
higher proper motion of about 40 to 50~mas\,yr$^{-1}$.
This trend continues for clouds at large separation from
$2''$ to about $75''$ or from 440~AU up to more than $10\,000$~AU
with pm-values in
the range of about $100-250~{\rm mas\,yr}^{-1}$. This finding agrees with
previous studies \citep{Paresce94,Maekinen04,Melnikov18,Liimets18}, 
which also found that the cloud motions are low for
$r\lapprox 2''$, while the extended H$\alpha$ nebula shows a much faster
systematic expansion.

We consider now the compact, fainter 
c-type H$\alpha$ clouds plotted in Fig.~\ref{DiagPMHalpha} with
green symbols. Several of these clouds stand out because they
show for their small separation $r\lapprox 1''$ a
significantly larger proper motion of about pm$'_r=30$ to 60~mas\,yr$^{-1}$,
when compared to pm$'_r=15$ to 25~mas\,yr$^{-1}$ 
for e-type and d-type clouds at similar separation.
These c-type clouds are transient, and they show no dust scattering signal
despite the fact that they are near to the illuminating Mira, so that
dust could produce potentially a detectable polarization. All these
properties are compatible with transient clouds formed by
instabilities in the collimated outflow. 

There are also slow moving c-type H$\alpha$ clouds, including
C$_{\rm NW}$, D6$_{\rm SW}$, D7$_{\rm SW}$, and K$_{\rm SW}$.
They could be just smaller clouds of the slow Mira wind
located in the wedge regions ionized by the hot component. 

\subsection{Interpretation for the H$\alpha$ outflow}

R~Aqr is a prototype system for an AGB binary with a bipolar
outflow and for a progenitor of an asymmetric planetary nebula.
The system parameters are typical for the wind Roche-lobe overflow
models producing the observed high contrast in the gas density
between the orbital plane and polar directions
\citep[e.g.,][]{Morris87,Soker94}.

The polar outflow producing a bipolar ionized
nebula is powered in many models by a jet or a strong disk wind from the
accreting companion. However, in R~Aqr there
are no features pointing to a powerful fast outflow with
$v_{\rm jet}>50~{\rm km\,s}^{-1}$ from the accreting star. Instead, the
main power source for the polar outflow is apparently
the substantial ionizing radiation with
$L_{\rm UV}\approx 10$ to $50~{\rm L}_\odot$ 
from the hot component. This produces a
strongly ionized, high density nebula in the center at
$r\lapprox 10$~AU. The bright H$\alpha$ clouds near the
star ($r\approx 20-200$~AU) move with a low velocity
of $\approx 20~{\rm km\,s}^{-1}$
and they look like irradiated clouds of the dust-driven Mira wind
often with dust scattering polarization.

The ionizing radiation escaping from the hot component is
apparently strongly collimated and this causes ionization
and heating of the Mira wind in two narrow, roughly polar wedges
with respect to the flattened wind density distribution aligned
with the orbital plane. The heating by the photoionization
produces most likely overpressured gas causing an
expansion in the direction of the polar density gradient, and this
could explain the observed acceleration of the H$\alpha$ clouds 
from $\approx 20~{\rm km\,s}^{-1}$ around 20~AU
to $\approx 100~{\rm km\,s}^{-1}$ around 400~AU. 
The radially moving large clouds expand with the falling ambient
pressure, the cloud opacities drop, and the initially
shielded clouds centers become transparent and are also
ionized and this could be responsible that the dust concentration
dissolve in the ionized outflow for separations $\gapprox 200$~AU. 

In this region the outflowing gas shows small, faster moving
transient clouds, possibly because of flow
instabilities similar to the clumpy winds of hot, luminous stars.
The observed turbulent flow produces certainly some cloud collisions
and shocks contributing to the heating of the gas in this region.
However, the photoionization from the central source is still the
main driver for the gas ionization, because
all H$\alpha$ clouds at $r\lapprox 400$~AU at $\theta>250^\circ$
react simultaneously to
the dust eclipse event from 2021 and the associated
overall shift of the illumination direction for the
SW region by more than 10$^\circ$. 

The outflowing gas shows at separations of $r\approx 1000$~AU
velocities up to 200~${\rm km\,s}^{-1}$, and for
$r\gapprox 1000$~AU it seems to expand
into the two polar bubbles, probably because the ionized gas is 
no more pressure-confined by the neutral Mira wind.
The bubbles and the extended equatorial ring of ionized gas
show a filamentary structures pointing to large
scale gas turbulence. In this region collisional ionization may be more
important than photoionization to keep the
gas ionized, at least for position angles not illuminated by
the collimated emission from the hot component.

This tentative scenario for the H$\alpha$ outflows
in R~Aqr is quite similar to the classical champagne flow model
of \cite{TenorioT79} for compact H{\sc II} regions breaking out of a dense
molecular cloud. Similar physical principles may also apply
for the two polar outflows of the R~Aqr system with its
strongly ionizing hot component in the center and the flattened cold gas
distribution.

According to this scenario there is no powerful
jet source in R~Aqr produced by an accreting star
with hydromagnetic acceleration of gas from the
fast rotating regions of the inner disk \citep[e.g.,][]{Livio97}.
This is confirmed by the H$\alpha$ clouds in the innermost
outflow region of R~Aqr,
which are just illuminated dusty clouds of the Mira wind.
The two polar outflows point to a hot companion emitting a
substantial amount of ionized radiation in two opposite directions,
probably because it is surrounded by a thick accretion torus
channeling the escaping radiation. This source moves on an
eccentric orbit inward and outward in the flattened,
dense neutral wind, perhaps even on an inclined orbit, and this creates
certainly various types of phase dependent effects. The most obvious
one is the dust obscuration event, but there are also variations
for the direction and strength of the collimated outflows and
many other complex and fascinating structures in the
circumstellar environment of R~Aqr, which warrant more
hydrodynamical studies based on high spatial resolution observations.

\section{Summary and discussions}
\label{Discussion}

The SPHERE/ZIMPOL high-resolution observations provide
for the R~Aqr system detailed maps for the H$\alpha$ emission of the ionized
gas in the collimated outflow and the scattering polarization 
of the dust in the Mira wind. These emission componenents are
ideal for the study of the binary interaction in this system. 
The multi-epoch data reveal the evolution of
individual features, and this provides
a much improved understanding of the complex radiation-hydrodynamic
taking place in R~Aqr.

\subsection{Binary motion and dust obscuration event}

The orbit of the hot component relative to the Mira
can be inferred from the differential H$\alpha$ imaging, which
shows a systematic motion of the central H$\alpha$ cloud
from west to east, confirming previous data. Near conjunction,
when the the hot component and the comoving dust cloud produce
an eclipse phase, the central
H$\alpha$ emission and the Mira star are practically aligned.
Because the continuum of the hot
star is not detected, one should expect small offsets
between the exact location of the star and the center of the
H$\alpha$ feature. Depite this, further monitoring
of the orbital motion of the hot component using the
central H$\alpha$ emission is important for obtaining tighter
constraints on the binary orbit and the stellar masses.

For the epoch 2021.5, during the dust eclipse event,
the $Q_\phi^{\rm ex}$ maps for the circumstellar polarization
show a clear signal reduction for the region inside about $r<10$~AU.
This event provides information about the 
size and extinction of the dust cloud moving with the
accreting component. 

\subsection{Circumstellar dust distribution and outflow motion}

From 10~AU to about 400~AU, the imaging polarimetry of R~Aqr shows
a radial decrease of the scattering polarization according to a power law of
$Q_\phi^{\rm ex}(r)\propto r^\alpha$ with an exponent of $\alpha=-2.7$. In
addition, $Q_\phi^{\rm ex}$ shows a flattened distribution orientated roughly
east-west, indicating an enhanced dust density near the inclined
orbital plane in agreement with the extended far-IR emission of
dust \citep{Mayer13}. We also observed a dust density step on the east side resembling
very much the models for the edge-on view of a spiral wind pattern
produced by an AGB-star binary \citep[e.g.,][]{KimH24}. 

The dust distribution in the
central region $r\lapprox 40$~AU shows strong temporal
changes during the covered 6.7~years. For the first three epochs,
there is about twice as much scattering polarization
on the eastern side of the Mira when compared to the west,
while the signal is about equal on both sides for the
dust eclipse epoch in 2021.5. Very close to the Mira, at about
$r\approx 10$ to 20~AU, dust bubbles were seen with sizes
and expansion velocities very similar to models of dust forming clouds
in AGB star winds \citep[e.g.,][]{Freytag23}.
The dust distribution and its evolution
in this region is also very similar to the CO gas observed with ALMA,
and some of the changes can be explained by dusty CO structures
corotating with the hot component in the orbital plane as
proposed by \cite{Bujarrabal21}.

The scattering polarization at separations from
10~AU to 300~AU shows about a dozen dust clouds producing a local maxima
in the $Q_\phi^{\rm ex}$ maps. 
Almost all dust clouds are present in all four 
epochs and show systematic radial proper motions. The derived
dust outflow velocity $v'_d\approx 21$~km/s
is significantly higher than the velocity
$|RV|_{\rm CO}\approx 17$~km/s for the molecular CO gas.
This yields a dust to gas drift velocity ratio of
$\Lambda=v'_d/|RV|_{\rm CO}\approx 1.3$
for the Mira wind in R~Aqr. 
This is a very interesting parameter for dust-driven winds, 
but the interpretation of this value must also consider the highly
asymmetric outflow pattern of the R~Aqr system.

\paragraph{Dust clouds in the ionized region.}
The hot star in R~Aqr produces a lot of ionizing
radiation and an ionized outflow. Dust clouds are
even detected in this region and the SPHERE/ZIMPOL capabilities
are ideal to determine the relative position of scattering dust
clouds and H$\alpha$ emission regions. We find 
small offset between polarization and H$\alpha$ signals,
because only the cloud surface facing the
central binary is ionized and the dust concentration in
the cloud core is shielded from ionization processes.
This type of dusty H$\alpha$ cloud are only seen for
separations closer than about 200~AU from the central binary.
For larger distances 
the dust concentrations seem to ``dissolve'' in the ionized outflow regions,
probably because the clouds expand with distance and become
transparent for the ionizing radiation.

There is a very prominent comet-like cloud in the
ionized outflow region with a dust core and dusty tail features.
The tail structures move faster away from the central binary than the core
as expected for dust particles accelerated by radiation pressure or
possibly drag forces of an ionized gas outflow.
We observe fan-like tail structures which could point to a
magneto-hydrodynamical effect. Another comet-like cloud
located in the neutral outflow region shows a straight dust tail
and no fan-like structure. The interaction with the hot component
could also have a strong impact on the chemistry of the 
circumstellar dust and gas \citep[e.g.,]{VandeSande22}, and 
R~Aqr is a unique laboratory for such investigations for AGB stars,
because one can study dust properties for clouds with and
without strong exposure to the far-UV radiation and the ionized outflow.

\subsection{H$\alpha$ emission clouds and ionized outflow}

The R~Aqr binary is famous for its collimated outflows of ionized gas
in the NE and SW direction and its extended nebulosity, which have been discussed
in many previous papers
\citep[e.g.,][]{Solf85,Paresce94,Liimets18,Melnikov18,Huang23}.
The SPHERE/ZIMPOL observations together with data from
HST show the evolution of the H$\alpha$ emission within
$2''$ or 450~AU in unprecedented detail and a dense temporal
coverage including the dust extinction phase around 2019 to 2023.
We describe in this paper the evolution
of the cloud structures in detail and derive their motion and life time.

The combination of H$\alpha$ and
polarization maps shows for the first time that
several prominent clouds near the central binary ($<1''$)
are just partly ionized gas clouds of the dusty Mira wind moving
with about $20~{\rm km\,s}^{-1}$. The presence of dust proves
that this gas does not originate from the accreting and
ionizing hot component. 
The prominent H$\alpha$ clouds
move steadily outward and expand with separation.

Within this region, there are also small
short-lived H$\alpha$ clouds that move with a higher
proper motion of about $50~{\rm km\,s}^{-1}$, instead of
the $20~{\rm km\,s}^{-1}$. These clouds
can be explained by instabilities in a clumpy outflow that is
perhaps similar to stellar winds of hot stars, but in R~Aqr the
clump hydrodynamics can be spatially resolved.

The typical proper motions of extended clouds increase
with separation and
reach for $r\approx 2''$ velocities of about $100~{\rm km\,s}^{-1}$
and up to $200~{\rm km\,s}^{-1}$ for $r>4''$. 
We suggest that the acceleration of the H$\alpha$ clouds
is caused by the overpressure in the narrow ionized outflow
channels, which is mainly heated by the photoionization through
the highly collimated energetic radiation from the hot component.
This scenario is strongly supported by the observed angular shift
in the distribution of emission line clouds in the SW taking
place from about 2018 to 2023 as well as the dust
obscuration event. This shift happens practically simultaneously
for cloud separations from of $0.3''$ (60~AU)
up to separations of $3''$ according to the VLT and HST data.

The scenario put forward for the two
H$\alpha$ outflows in R~Aqr is quite similar to the classical champagne
flow model of \cite{TenorioT79} for a compact
H\,{\sc ii} region breaking out of a dense molecular cloud.
Similar physical principles may apply for the polar outflows in the
R~Aqr system with its flattened cold gas distribution created by
the interaction of the strong Mira wind with the stellar companion.

\subsection{R~Aqr in the context of other AGB star binaries}
The symbiotic Mira R~Aqr is
a prototype system for the formation of bipolar planetary nebulae that form as a
result of the mass loss from an AGB star binary
\citep[e.g.,][]{Corradi95}.
He2-104 or BI Cru are two other well-known symbiotic Miras
with a similar bipolar nebulae \citep[e.g.,][]{Corradi93}, and
HM Sge, V1016 Cyg, and Hen2-147 at least have extended nebular structures
\citep{Corradi99,SantanderG07}. 
R~Aqr is outstanding within this group because it is at
least ten times closer, and the central binary can be spatially resolved
with modern imaging techniques. R~Aqr is also special because
the luminosity of $\approx 10-50~{\rm L}_\odot$ for the hot component
is about a hundred times smaller than for the other symbiotic Miras
with very bright emission nebulae. However, the luminosity of R~Aqr B
is still large enough to create substantial ionization effects within
the dusty wind of the Mira star and a strongly reduced abundance
of CO molecules \citep{Bujarrabal21}, and it
possibly also impacts the formation and evolution of dust
particles in the Mira wind \citep[e.g.,][]{VandeSande24}. 
In addition, the overall geometry of the dusty Mira wind
appears to be similar to the models for an AGB binary in the
wind Roche-lobe overflow 
regime \citep[e.g.,][]{Mastrodemos99,Mohamed12}.
The SPHERE/ZIMPOL and ALMA maps show a Mira wind with the expected 
higher dust and gas density in the equatorial plane
probably because of the outflow focussing by the companion, the
influence of the orbital motion, and possibly the rotational spin-up
of the AGB star together with an ionized polar outflow originating
from the accreting companion.

R~Aqr is therefore a very interesting object within the very diverse group
of resolved, nearby AGB star binaries. Other members of this group
are, for example, the Mira A and B system \citep{Khouri26}, V Hya
\citep{Planquart24}, and $\pi^1$~Gru A with its companion \citep{Montarges25}, as all have binary separations in the range from 10~AU to 100~AU. Within
this group, only R~Aqr shows substantial ionization, providing
a special aspect when compared to the other systems
with less powerful companions. The combination of the results for all of these
objects are required for a better understanding of
the complex hydrodynamics in AGB star binaries
\citep[see, e.g.,][]{Decin21}, and R~Aqr is a key object for the
investigation of the impact of an ionizing companion
on the dusty Mira wind.

\section*{Data availability}
The reduced count maps of the differential H$\alpha$ emission 
$I_{{\rm H}\alpha}$ and the I-band, zp-corrected, azimuthal polarization
$Q_\phi^{\rm ex}$ are available for all four epochs in electronic form
at the CDS via anonymous ftp to cdsarc.u-strasbg.fr (130.79.128.5) or via
http://cdsweb.u-strasbg.fr/cgi-bin/qcat?J/A+A/.
All SPHERE/ZIMPOL images in this paper are based on these maps.

\begin{acknowledgements}
SPHERE is an instrument designed and built by a consortium consisting
of IPAG (Grenoble, France), MPIA (Heidelberg, Germany), LAM (Marseille,
France), LESIA (Paris, France), Laboratoire Lagrange (Nice, France), 
INAF – Osservatorio di Padova (Italy), Observatoire de Gen\`eve 
(Switzerland), ETH Zurich (Switzerland), NOVA (Netherlands), ONERA (France) 
and ASTRON (Netherlands), in collaboration with ESO. 
SPHERE was funded by ESO, with additional contributions from CNRS (France), 
MPIA (Germany), INAF (Italy), FINES (Switzerland) and NOVA (Netherlands). 
SPHERE also received funding from the European Commission Sixth and 
Seventh Framework Programmes as part of the Optical Infrared 
Coordination Network for Astronomy (OPTICON)
under grant number RII3-Ct-2004-001566 for FP6 (2004–2008), grant number
226604 for FP7 (2009–2012) and grant number 312430 for FP7 (2013–2016).

We acknowledge with thanks the variable star observations
from the AAVSO International Database contributed by observers worldwide
and used in this research.
\end{acknowledgements}

\bibliographystyle{aa} 
\bibliography{aa61903-26corr} 

\vfil\eject
\clearpage

\begin{appendix}

\section{Maps for the epoch e$_2$, e$_3$, and e$_4$}
\label{Maps234}

  \begin{figure}[htbp]
\includegraphics[trim=0.cm 0.cm 1.5cm 1.5cm,clip,width=8.5cm]{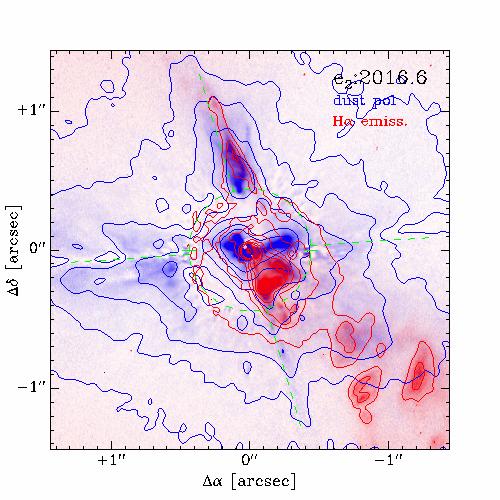}  
\caption{H$\alpha$ intensity $I_{{\rm H}\alpha}$ (red) and I-band 
  polarization $Q_{\phi}^{\rm ex}$ (blue) for the R~Aqr observations
  from 2016. The contours are smoothed and given
  in increments of a factor of two for the $I_{{\rm H}\alpha}$ and a factor of 3.16
  for $Q_{\phi}^{\rm ex}$. The colors are scaled with separation
  $\propto r^\alpha$ with $\alpha\approx 2$ to make
  the much weaker emission at larger $r$ visible.
  Dashed green lines mark regions affected by PSF features.}
\label{HaDust16}
\end{figure}

\begin{figure}
\includegraphics[trim=0.cm 0.cm 1.5cm 1.5cm,clip,width=8.5cm]{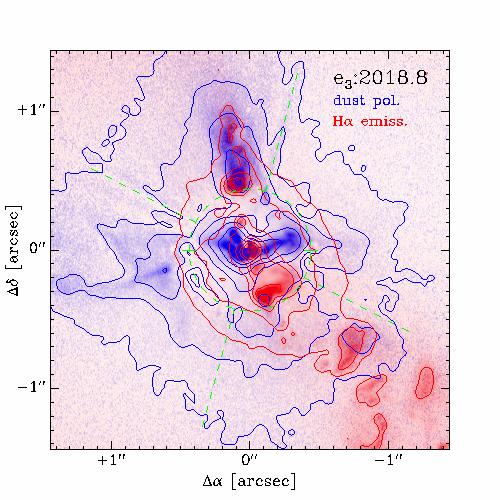}  
\caption{Same as Fig.~\ref{HaDust16} but for the observations from
 2018.}
\label{HaDust18}
\end{figure}

\begin{figure}
\includegraphics[trim=0.cm 0.cm 1.5cm 1.5cm,clip,width=8.5cm]{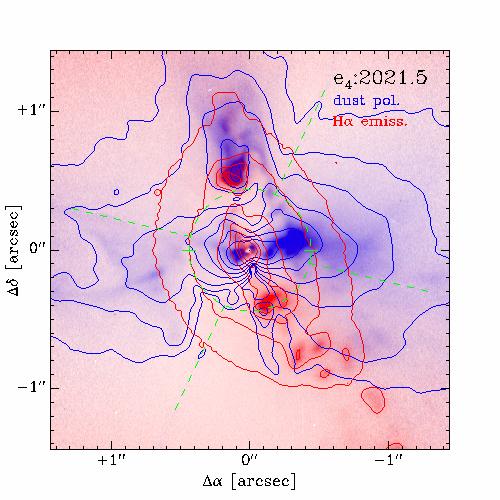}  
\caption{Same as Fig.~\ref{HaDust16} but for the observations from
 2021.}
\label{HaDust21}
\end{figure}

\noindent
The maps in the Figs.~\ref{HaDust16}, \ref{HaDust18}, and \ref{HaDust21}
illustrate for the epochs e$_2$, e$_3$, and e$_4$
the distributions of the signals for the H$\alpha$ emission
and the scattering polarization.
These figures complement the map in Fig.~\ref{HaDust14}
for the first epoch. The temporal changes in the overall
signal distribution can be inferred from the contours, while the
color scales show the evolution of small scale features. The smoothed
contours for the H$\alpha$ emission are given for $I_{{\rm H}\alpha}$ 
count values $[20,40,80,160,320,640]\cdot c$
covering a range of a factor of 32 for the selected
50~s integrations. The scale parameter is $c=1$ for the first
three epochs and $c=0.5$ for the last epoch e$_4$ to account for the
reduced AO performance of the last epoch. The contours for
the $Q_\phi^{\rm ex}$ polarization are plotted for values
$[0.3,1,3,10,30,100,300]\cdot c$ 
covering a factor of 1000. The
parameter $c$ was selected, so that the peak pixel of the
bright dust cloud about $0.4''$ north of the center
has a signal of about 60 in arbitrary units.
The color scaling uses radial power laws, such as
$r^{-\alpha}$ with $\alpha\approx 2$, 
and small offset parameters for large and small radii. An individual
fine tuning of the red and blue color scales was applied for each
map to reveal better small scale structures at different separations.
When using only a simple radial power law for the color scaling then
many structures would be hardly visible. Therefore, one should rely on the
contours for estimating the surface brightness of nebular structures
in the maps. 

The H$\alpha$ maps show many discrete clouds and an extended halo.
Much of this halo is produced by the extended PSF of the 
instrument and the presence of a possible
diffuse H$\alpha$ emission is uncertain (see Paper I). The PSF smearing
also affects the polarization signal in a complex way as described
in \cite{Schmid26}. Therefore, the extended, diffuse signal
in the presented maps should be interpreted with caution.

\begin{table}
  \caption{R~Aqr data from SPHERE/ZIMPOL selected for this study.}
  \label{DataTab}
\noindent
\begin{tabular}{llccccccc}
  \noalign{\smallskip\hrule\smallskip}
$\hspace{-2.5mm}$
filters & mod.  & DIT & nDIT & coro & peak & \multispan{2}{\hfil region \hfil} \\
        & mode  &  [s]&      & ND & [kcts]    & in & out \\ 
\noalign{\smallskip\hrule\smallskip}

\noalign{\smallskip \noindent epoch e$_1$: 2014.8 \smallskip}

\noalign{\smallskip}
$\hspace{-2.5mm}$
Cnt/Ha\tablefootmark{a}  & slow    & 50  & 2        
                               &       &  19/7 & $\times$ & $\times$ \\ 
\noalign{\smallskip}
$\hspace{-2.5mm}$
Cnt820    & fast    & 1.2 & 10   
                          &        &  22   & $\times$  & \\  
$\hspace{-2.5mm}$
I\_PRIM   & fast    & 1.2 & 10   
                          &        &  sat   & $\times$  &  \\ 
$\hspace{-2.5mm}$
I\_PRIM & slow   & 10  & 20   
                          & coro   &  --  &    &  $\times$  \\ 
\noalign{\smallskip \noindent epoch e$_2$: 2016.6 \smallskip}

\noalign{\smallskip}
$\hspace{-2.5mm}$
Cnt/Ha & slow  & 10  & 2    
&        & $\hspace{-2mm}65\tablefootmark{b}/16\hspace{-2mm}$ &   $\times$   &     \\
$\hspace{-2.5mm}$
Cnt/Ha & slow  & 50  & 2    
                           &        &  sat/sat & ($\times$) &  $\times$ \\    
\noalign{\smallskip}
$\hspace{-2.5mm}$
Cnt820  & fast   & 1.2  & 6        
                           &       &    38   &  $\times$ &    \\
$\hspace{-2.5mm}$
N\_I  & fast    & 1.2  & 6        
                           & ND1      &   16    & $\times$  &   \\
$\hspace{-2.5mm}$
N\_I  & fast        & 1.2  & 6       
                           &         &  sat     &  $(\times)$  &    \\
$\hspace{-2.5mm}$
N\_I   &  slow  & 10.  & 6        
                            &        &  sat   &     & $\times$ \\
$\hspace{-2.5mm}$
N\_I  & slow    & 10.  & 20       
                           & coro   &    --   &     & $\times$  \\
\noalign{\smallskip \noindent epoch e$_3$: 2018.8 \smallskip}

\noalign{\smallskip}
$\hspace{-2.5mm}$
Cnt/Ha & slow  & 50  & 2       
&             &  $\hspace{-2mm}63/23\hspace{-2mm}$
                                                  & $\times$  & $\times$ \\
\noalign{\smallskip}
$\hspace{-2.5mm}$
N\_I  & fast    & 1.2  & 10        
                          &          &   60   & $\times$    &            \\
$\hspace{-2.5mm}$
N\_I  & slow   & 10  & 10        
                          & coro     &  -- &   & $\times$ \\
$\hspace{-2.5mm}$
N\_I  & slow  & 10  & 10        
                          & coro     &  -- &   & $\times$ \\
\noalign{\smallskip \noindent epoch e$_4$: 2021.5 \smallskip}

\noalign{\smallskip}
$\hspace{-2.5mm}$
Cnt/Ha & slow  & 10  & 4           
                           &             &  5/1  & ($\times$) &   \\
$\hspace{-2.5mm}$
Cnt/Ha & slow  & 50  & 2        
                           &             &  $\hspace{-2mm}54/12\hspace{-2mm}$ & $\times$ & $\times$  \\
\noalign{\smallskip}
$\hspace{-2.5mm}$
N\_I     & fast     & 5.0  & 20        
                          &          &  sat    & $\times$    &    \\
$\hspace{-2.5mm}$
N\_I    & fast    & 2.0  & 50        
                           &         &  31  & $\times$    &  \\
$\hspace{-2.5mm}$
N\_I     & slow  & 10  & 10        
                          &         &  sat  &     &   $\times$   \\
$\hspace{-2.5mm}$
N\_I    & slow    & 10  & 10      
                          &         &  sat  &     &   $\times$   \\
\noalign{\smallskip}
\noalign{\smallskip\hrule}
\end{tabular}
\tablefoot{The columns list the used filters,
the modulation mode, detector integration times (DIT) in seconds,
number of integrations (nDIT) per HWP position angle;
\tablefoottext{a}{short for filter CntHa in arm~1 and filter N\_Ha in
arm~2;}
\tablefoottext{b}{a few pixels saturated.}}
\end{table}

\balance
  
\section{ZIMPOL data}
\label{ZIMPOLdata}

The SPHERE/ZIMPOL data used for
this work are listed in Table~\ref{DataTab}.
All observations were taken in
field stabilized mode using polarimetry. A polarimetric
dataset consists of a cycle with four exposures where
one exposure includes nDIT frames with a given detector integration
time (DIT). The four exposures are taken with different
half-wave plate orientations measuring the intensity of the
linear polarization for the directions $0^\circ$, $45^\circ$, $90^\circ$,
and $135^\circ$, which are then combined with a polarization switch and
a double difference method to derive the Stokes parameters
$Q = I_{0}-I_{90}$ and $U=I_{45}-I_{135}$ and the
corresponding intensities $I_Q=I_{0}+I_{90}$ and $I_U=I_{45}+I_{135}$
\citep{Schmid18}. ZIMPOL has
two arms and the polarization can be measured simultaneously
in two bands if different filters are inserted. For our
I-band polarimetry the same filter type was inserted
in both arms while an H$\alpha$ and a continuum filter are used
for simultaneous differential H$\alpha$ imaging (Table~\ref{DataTab}).
ZIMPOL uses a modulation
technique to measure the polarization and two modulation modes are
available: fast modulation for high flux observations with
short DITs and high detector gain but also higher read-out noise,
and a slow modulation for low flux observations with integrations
$\geq 10$~s and low detector gain providing low read-out noise.
We used fast modulation for the strong polarization signal of the
central star and the bright scattering region near the star, and
slow polarimetry for the much fainter outer scattering region
and the differential imaging in the narrow H$\alpha$ and continuum
H$\alpha$ filters. The $\times$-symbols in the two last columns indicate
the covered region and $(\times)$ mark datasets which
complement data with higher quality.

\paragraph{Filters.} For H$\alpha$ imaging the continuum
filter CntHa ($\lambda_C=644.9$~nm, $\Delta\lambda=4.1$~nm)
in ZIMPOL arm~1 and the narrow band filter N\_Ha
($\lambda_C=656.53$~nm and $\Delta\lambda=0.97$~nm) in arm~2
are used.
These data were taken in slow modulation mode with a
low gain and low read out noise.
Exposure times were selected to avoid saturation
by the Mira variable in the CntHa filter. In 2016 the Mira was very bright
and saturated a few pixels, even in the short 10~s integration.
Better differential maps $I_{{\rm H}\alpha}$ are obtained if only high
quality pairs of N\_Ha and CntHa are selected for the continuum subtraction.

For the I-band polarimetry
short integration with fast modulation mode and the filter Cnt820
($\lambda_C=817.3$~nm, $\Delta\lambda=19.8$~nm) or
the filters N\_I ($\lambda_C=817$~nm, $\Delta\lambda=81$~nm)
were used for the inner scattering region. For epoch e$_2$ with
bright Mira the combination of N\_I filters
with the neutral density filter ND1 with an attenuation of about a
factor of 11 were inserted according to the VLT/SPHERE user manual.

The outer scattering region was obtained with the
N\_I filter or filter
I\_PRIM ($\lambda_C=790$~nm, $\Delta\lambda=153$~nm) with
DITs of 10~s or longer in slow modulation mode. In this case the
central region was either covered by a coronagraph or the detector
did saturate, but without producing harmful effects
outside the saturation region. The used coronagraphic focal plane mask
has a radius of 77.5~mas. The column ``peak'' in
Table~\ref{DataTab} gives for
non-coronagraphic data the peak counts at the position of
the Mira. The flag ``sat'' is given if the detector saturated for
a level $> 64$~k counts. 
Stokes $Q$ and $U$ maps for the I-band polarimetry of epoch e$_1$
are described in detail in \cite{Schmid18}.

\vfill\eject

\end{appendix}

\end{document}